\documentclass[10pt,letterpaper]{article}

\usepackage{chuancmds}

\usepackage[top=0.85in,left=1.5in,footskip=0.75in]{geometry}

\usepackage{changepage}

\usepackage[utf8]{inputenc}

\usepackage{textcomp,marvosym}

\usepackage{fixltx2e}

\usepackage{amsmath,amssymb}

\usepackage{cite}

\usepackage{nameref,hyperref}

\usepackage{microtype}
\DisableLigatures[f]{encoding = *, family = * }

\usepackage{rotating}

\raggedright
\usepackage[aboveskip=1pt,labelfont=bf,labelsep=period,justification=raggedright,singlelinecheck=off]{caption}

\makeatletter
\renewcommand{\@biblabel}[1]{\quad#1.}
\makeatother

\date{}

\newcommand{\tskip}{\vspace*{0cm}}
\newcommand{\micron}{\mu \rm{m}}

\newcommand{\rms}{\rm{s}}
\newcommand{\hour}{\,\rm{h}}
\newcommand{\pN}{\rm{pN}}

\newcommand{\void}[1]{}

\begin{document}
\vspace*{0.35in}

\begin{flushleft}

{\Large
\textbf\newline{A Stochastic Multiscale Model that Explains the Segregation of Axonal Microtubules and Neurofilaments 
 in Neurological Diseases}
}
\newline
\\
Chuan Xue\textsuperscript{1,*},
Blerta Shtylla\textsuperscript{2},
Anthony Brown\textsuperscript{3} 
\\
\bigskip
{\bf{1}} Department of Mathematics, Ohio State University, Columbus, Ohio, United States of America\\
{\bf{2}} Department of Mathematics, Pomona College,  Claremont, California, United States of America\\
{\bf{3}} Department of Neuroscience, Ohio State University,  Columbus, Ohio, United States of America\\
\bigskip

* cxue@math.osu.edu

\end{flushleft}

\section*{Abstract}
The organization of the axonal cytoskeleton is a key determinant of the normal function of an axon, which is a long thin projection away from a neuron. Under normal conditions two axonal cytoskeletal polymers microtubules and neurofilaments align longitudinally in axons and are interspersed in axonal cross-sections. However, in many neurotoxic and neurodegenerative disorders, microtubules and neurofilaments segregate apart from each other, with microtubules and membranous organelles clustered centrally and neurofilaments displaced to the periphery. This striking segregation precedes abnormal and excessive neurofilament accumulation in these diseases, which in turn leads to focal axonal swellings. While neurofilament accumulation suggests the impairment of neurofilament transport along axons, the underlying mechanism of their segregation from microtubules remains poorly understood for over 30 years.To address this question, we developed a stochastic multiscale model for the cross-sectional distribution of microtubules and neurofilaments in axons.  The model describes microtubules, neurofilaments and organelles as interacting particles in a 2D cross-section, and is built upon molecular processes that occur on a time scale of seconds or shorter.  It incorporates the longitudinal transport of neurofilaments and organelles through this domain by allowing stochastic arrival and departure of these cargoes, and integrates the dynamic interactions of these cargoes with microtubules mediated by molecular motors. Simulations of the model demonstrate that organelles can pull nearby microtubules together, and in the absence of neurofilament transport, this mechanism gradually segregates microtubules from neurofilaments on a time scale of hours, similar to that observed in toxic neuropathies. This suggests that the microtubule-neurofilament segregation is simply a consequence of the selective impairment of neurofilament transport.  The model generates the experimentally testable prediction that the rate and extent of segregation will be dependent on the sizes of the moving organelles as well as the density of their traffic.
 
\section*{Author Summary}
The shape and function of axons is dependent on a dynamic system of microscopic intracellular protein polymers (microtubules, neurofilaments and microfilaments) that comprise the axonal cytoskeleton.  Neurofilaments are cargoes of intracellular transport that move along microtubule tracks, and they accumulate abnormally in axons in many neurotoxic and neurodegenerative disorders. Intriguingly, it has been reported that neurofilaments and microtubules, which are normally interspersed in axonal cross-sections, often segregate apart from each other in these disorders, which is something that is never observed in healthy axons. Here we describe a stochastic multiscale computational model that explains the mechanism of this striking segregation and offers insights into the mechanism of neurofilament accumulation in disease.


\section*{Introduction}
 
Axons are long slender projections of nerve cells that permit fast and specific electrical communication with other cells over long distances.  The ability of nerve cells to extend and maintain these processes is critically dependent on the cytoskeleton, which is a dynamic scaffold of microscopic protein polymers found in the cytoplasm of all eukaryotic cells.  The axonal cytoskeleton comprises microtubules, intermediate filaments called neurofilaments, and microfilaments.  Microtubules and neurofilaments are both long polymers that align in parallel along the long axis of the axon, forming a continuous overlapping array that extends from the cell body to the axon tip \cite{Tsukita1981,Schnapp1982}. Microtubules are stiff hollow cylindrical structures about 25 nm in diameter that serve as tracks for the long-range bidirectional movement of intracellular membranous organelles and macromolecular cargo complexes.  In axons, this movement is known as axonal transport \cite{Brown2013}.  The cargoes of axonal transport are conveyed by microtubule motor proteins: kinesins in the anterograde direction (towards the axon tip), and dyneins in the retrograde direction (towards the cell body) \cite{Hirokawa2010}.  Neurofilaments, which are the intermediate filaments of nerve cells, are flexible rope-like polymers that measure about 10 nm in diameter \cite{Perrot 2009}.  These polymers function as space-filling structures that expand axonal cross-sectional area, thereby maximizing the rate of propagation of the nerve impulse \cite{Cleveland1991,Hoffman1995}. In large axons, neurofilaments are the single most abundant structure and occupy most of the axonal volume \cite{Friede1970}.   Mutant animals that lack neurofilaments develop smaller caliber axons and exhibit delayed conduction velocities \cite{Sakaguchi1993,Eyer1994,Zhu1997}.  

In addition to their structural role in axons, neurofilaments are also cargoes of axonal transport, moving along microtubule tracks powered by kinesin and dynein motors \cite{Wang2000,Xia2003,Francis2005,He2005,Uchida2009}.  The filaments move at rates similar to membranous organelles but the movements are less frequent, resulting in a ``stop and go'' motile behavior characterized by short bouts of movement interrupted by prolonged pauses on a time scale of seconds or shorter \cite{Brown2009SAT,Brown2003d}.  The net result is an average rate of transport that is much slower than that for many other cargoes.  

Neurofilaments has been observed to accumulate abnormally in axons in many neurodegenerative diseases including amyotrophic lateral sclerosis, hereditary spastic paraplegia, giant axonal neuropathy and Charcot-Marie-Tooth disease (also known as hereditary distal motor and sensory neuropathy) \cite{Al-Chalabi2003,Rao2003a,Liu2004,Lariviere2004b,Perrot2009,Liem2009}, and also in many toxic neuropathies\cite{Griffin1983,Sayre1985,Gold1987,Graham1999,Llorens2013}. In extreme cases, these accumulations can lead to giant balloon-like axonal swellings \cite{Delisle1984,Hirano1984,Sasaki1990,Sasaki1992,Fabrizi2004,Fabrizi2007}.  These accumulations are thought to be caused by alterations in neurofilament transport, but the mechanism is not understood \cite{Julien1995,Collard1995,Julien1997,Miller2002,DeVos2008,Millecamps2013}. 

\begin{figure}[h]
\begin{center}
\includegraphics{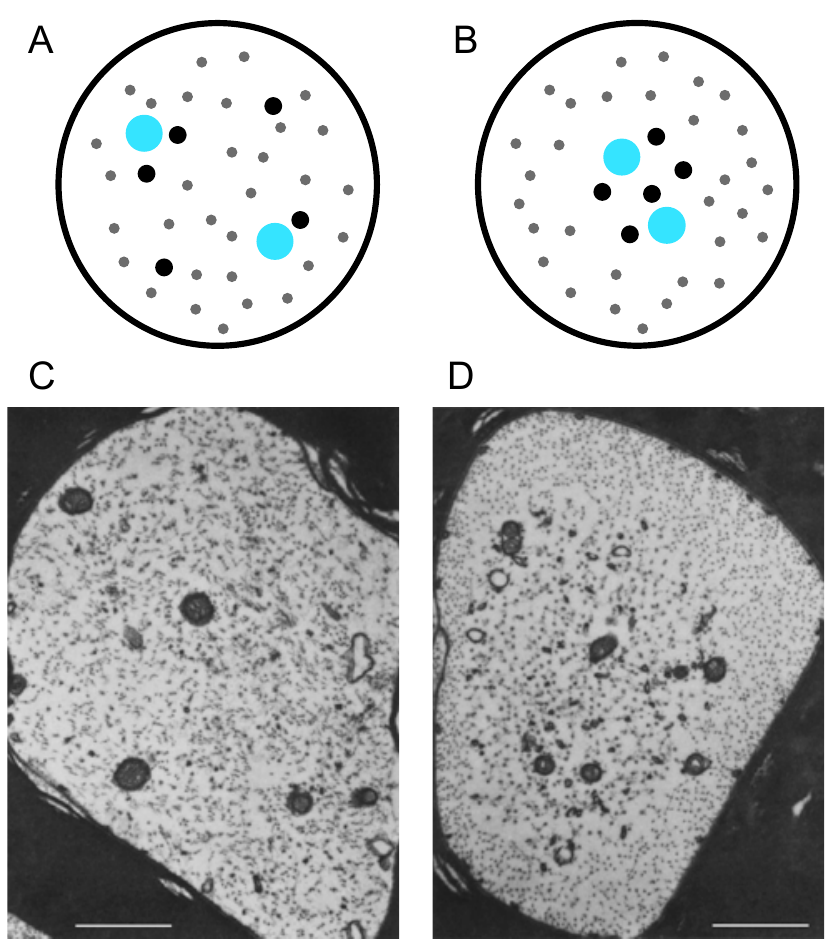}
\end{center}
\caption{ { Cross-sectional distributions of microtubules and neurofilaments via electron microscopy. } (A) Schematic drawing that illustrates the normal distribution of microtubules (large black dots), neurofilaments (small grey dots) and organelles (cyan disks) in untreated axons. (B) Schematic drawing that illustrates the segregated components in IDPN-treated axons.  (C) In normal axons  microtubules, neurofilaments and organelles are interspersed. Big dots: microtubules; small dots: neurofilaments; circular objects: organelles. (D) In IDPN-treated axons these components segregate in a core-tube pattern, with microtubules and organelles typically in the center and neurofilaments displaced to the periphery. The scale bars are 1 $\mu \rm{m}$. Reproduced from Papasozomenos et al, 1981. Originally published in Journal of Cell Biology, 91:866-871.} 
\label{fig_exp}
\end{figure}

In healthy axons, microtubules and neurofilaments align along the length of an axon and are interspersed in axonal cross-sections \cite{Price1988, Price1993,Tsukita1981,Hsieh1994}, with microtubules often forming small clusters in the vicinity of membranous organelles \cite{Price1991,Friede1970,Hirokawa1982}.   However, in many toxic and neurodegenerative disorders these polymers segregate, with microtubules and membranous organelles typically clustered in the center of the axon, and  neurofilaments displaced to the periphery (Fig. \ref{fig_exp}).   This striking cytoskeletal reorganization, which is never observed in healthy axons, has been reported in neurodegenerative disorders as diverse as giant axonal neuropathy \cite{Taratuto1990,Donaghy1988a,Griffiths1980} and Charcot-Marie-Tooth disease \cite{Goebel1986,Fabrizi2007}, as well as in neurotoxic neuropathies induced by exposure to agents as diverse as 2,5-hexanedione and 3,3'-iminodiproprionitrile  (IDPN)\cite{Papasozomenos1981,Papasozomenos1982,Griffin1983,Griffin1983CDI,Griffin1983b, Papasozomenos1985,Hirokawa1985,Papasozomenos1986,Nagele1988a}, aluminum \cite{Bizzi1984}, carbon disulfide \cite{Gottfried1985,Jirmanova1984}, estramustine phosphate \cite{Sahenk1992}, 1,2-diacetylbenzene \cite{Tshala-Katumbay2005} and 1,2,4-triethylbenzene \cite{Tshala-Katumbay2006}, and in a transgenic mouse expressing a mutant neurofilament protein \cite{Lee1994}.  However, the mechanism of this segregation and its relationship to the neurofilament accumulation that also occurs in these different conditions is not known. 

Microtubule-neurofilament segregation has been studied most extensively for IDPN and 2,5-hexanedione.  IDPN is a compound closely related to the naturally occurring food poison 3-aminopropionitrile which causes the neurological disorder lathyrism \cite{Selye:1957, Cadet1989a, Spencer1983, Spencer1991}, and 2,5-hexanedione is a metabolite of the industrial solvent hexane. The mechanism of toxicity is not known, but it is thought to involve  chemical modification of neurofilaments, which presumably disrupts their normal interactions with microtubules in some way\cite{Sayre1985,Morandi1987,Jacobson1987, Eyer1989,Mitsuishi1993,Zhu1998,Llorens2011,Llorens2013}. Systemic administration of IDPN or 2,5-hexanedione to rats causes selective impairment of neurofilament transport \cite{Griffin1978,Yokoyama1980,Griffin1985,Parhad1986,Komiya1986}, focal accumulations of axonal neurofilaments leading to axon enlargement, and neurological defects similar to amyotrophic lateral sclerosis (ALS) in humans \cite{Chou1964,Gold1986,Llorens2009,Clark1980}.   Injection of IDPN or 2,5-hexanedione into peripheral nerves results in local microtubule-neurofilament segregation within just a few hours, preceding the accumulation of neurofilaments by hours or days \cite{Papasozomenos1981, Papasozomenos1982, Griffin1982, Papasozomenos1986,Griffin1983CDI}. This segregation does not appear to affect the axonal transport of membranous organelles, which continue to interact with and move along these tracks in spite of their clustering. Moreover, in the case of IDPN the segregation has been shown to be reversible \cite{Papasozomenos1981, Griffin1983}, as has the impairment of neurofilament transport \cite{Komiya1987}. In  \cite{Griffin1983}, a single injection of IDPN into rat sciatic nerves resulted in segregation in axons at the injection site within a few hours, but the segregation disappeared in about a day. In \cite{Papasozomenos1981}, a single injection of IDPN into the body cavity of rats resulted in segregation within the axons of the sciatic nerve after 4 days, and this disappeared after six weeks. Thus the microtubule-neurofilament segregation caused by IDPN and 2,5-hexanedione is fast, local and reversible.
 
Though the segregation of microtubules and neurofilaments in axons was first described more than 30 years ago, the underlying mechanisms are still poorly understood. Given that neurofilaments move along microtubule tracks and that microtubule-neurofilament segregation precedes neurofilament accumulation and axonal enlargement in rodent models, it is attractive to speculate that the segregation reflects an uncoupling of neurofilaments from their transport machinery \cite{Griffin1978}. 
However, the mechanism by which such an uncoupling at the molecular level might generate polymer segregation at the population level remains unclear.
 
To address these questions, we have developed a stochastic multiscale model for the cross-sectional organization of microtubules and neurofilaments in axons. The model describes  microtubules, neurofilaments, and organelles   as interacting particles that move in a 2D domain representing a cross-section of an axon, and incorporates axonal transports of neurofilaments and organelles, as well as volume exclusion and Brownian motion of all the particles. Neurofilaments and organelles dynamically bind to and unbind from nearby microtubules through molecular motors, and the motor cross-bridges are modeled as elastic springs. The longitudinal movement of neurofilaments and organelles along axons is modeled by stochastic addition and removal of these cargoes. The multiscale nature of the model lies in that it is built upon molecular processes that occur on a time scale of seconds or fractions of a second, and addresses segregation phenomena of two populations of polymers  that occur on a time scale of hours to days.

Simulations of the model demonstrate that if we block neurofilament transport by preventing neurofilament from binding to microtubules, then organelles pull nearby microtubules together and gradually segregate them from neurofilaments, on the same time course as observed in toxic neuropathies; while if we restore neurofilament transport, then microtubules and neurofilaments start to remix until their spatial distribution returns to normal. The model further predicts that (1) during the segregation process, microtubules first form small clusters, small clusters merge into bigger clusters, and eventually a single cluster forms close to the center of the domain; (2) in the absence of neurofilament transport, larger organelles are more effective in causing complete cytoskeletal segregation than small organelles with the same density.   
 
\section*{Model}

\subsection*{The stochastic multiscale model}
\begin{figure}[h!]
\begin{center}
 \includegraphics{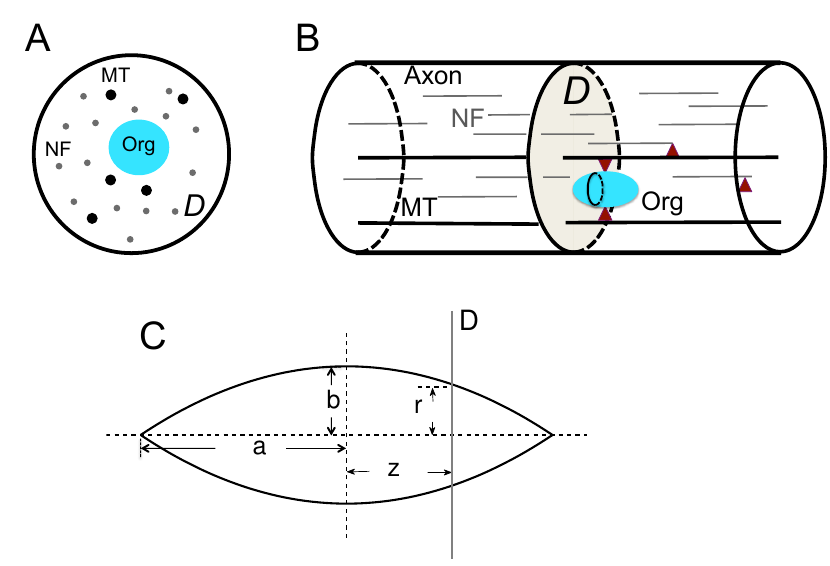}
\end{center}
\caption{ { Model setup.} (A) The model geometry. The computational domain $D$ represents a cross-section of an axon. The black circle is the domain boundary  representing the axon membrane. Small grey dots are neurofilaments (NF), large black dots are microtubules (MT) and cyan filled disks are organelles (Org). (B) The relation of $D$ to the whole axon. Thin grey lines are neurofilaments, thick black lines are microtubules, the cyan body is an organelle, red triangles represent molecular motors that move microtubules and organelles along microtubule tracks.   (C) The shape of  organelles considered in this model. The cross-sectional radius of an organelle in $D$ depends on its position relative to $D$. }\label{fig_geometry} \end{figure}

In our model,  microtubules, neurofilaments and organelles are described as individual particles that move in a circular domain $D$ with fixed radius $R_0$, representing a cross-section of an axon.  Microtubules and neurofilaments are rod-like polymers that align along the length of axons, thus they are treated as nondeformable disks in  $D$ (Fig. \ref{fig_geometry}A), with center positions denoted by $\bx^k_i = (x^k_i, y^k_i)$ and radii by $r_i^k$. Here $k = M$ or $N$  is the index for particle type: $M$ for microtubule,  $N$ for neurofilament;  and $i$ with $0 \leq i \leq n^k$ is the index for the $k$-type particle where $n^k$ is the total number of  $k$-type particles. The radii of microtubules and neurofilaments are constant, with $r_i^M = 12.5\, \rm{nm}$ and $r_i^N = 5\, \rm{nm}$. Organelles in axons have different sizes and shapes, and their cross-sectional geometry depends on their position relative  to the cross-section (Fig. \ref{fig_geometry}B). In this model, we took organelles as spindle-shaped objects and, for simplicity, we did not consider possible shape changes (Fig. \ref{fig_geometry}C). Therefore the organelles exist as non-deformable disks in $D$,  and as an organelle crosses $D$, its cross-sectional radius, $r^O_i$,  varies according to its position, $z_i^O$, relative to $D$, 
\beq\label{eqn_radius_FC}
r^O_i = b \left(1-\frac{(z^O_i)^2}{a^2}\right), \qquad  -a \leq z^O_i \leq a. 
\eeq
Here $a$ is half of the organelle length, $b$ is its maximum cross-sectional radius,  $z_i^O$ is the distance of its center to $D$, and the index ``O'' stands for organelle. By varying the parameters $a$ and $b$, we can vary the overall dimension of the organelles  
(Fig. \ref{fig_geometry}C). 

We examined three key molecular mechanisms that contribute to the cross-sectional distribution of microtubules and neurofilaments: slow axonal transport of neurofilaments, fast axonal transport of organelles, and volume exclusion of all the particles. In the following sections we describe in detail how these mechanisms were incorporated into our model. We denote the unit vector pointing from $\bx^{k}_i$ to $\bx^{l}_j$
by $ \bolde^{kl}_{ij}$, and the surface distance between the $i$-th particle of $k$-type and $j$-th particle of $l$-type by  $d^{kl}_{ij}$, given as, 
\beq
d^{kl}_{ij} = | \bx_i^{k } - \bx_j^{l} | - r_i^{k } - r_j^{l}. 
\eeq 

\subsubsection*{Mechanism 1: Slow axonal transport of neurofilaments.  } 
Neurofilaments interact with molecular motors (kinesin and dynein) which move these polymers along microtubules either anterogradely or retrogradely \cite{Uchida2009,Wagner2004,Francis2005,Shah2000}.  The movements are fast but infrequent because the filaments spend most of their time pausing, which results in a slow average rate of transport \cite{Wang2000,Trivedi2007}. This longitudinal movement of neurofilaments along microtubules can change the spatial distribution of these polymers in axonal cross-sections.  For example, if a neurofilament moves into the cross-section along a microtubule in a small cluster of microtubules it can displace one or more of the adjacent microtubules, dispersing the cluster.  Alternatively, if a neurofilament between two microtubules moves out of the cross-section, then the two microtubules are able to diffuse closer together.  

Based on the above considerations, we modeled neurofilament transport in the following way.  A neurofilament in $D$ can bind to a microtubule within a binding radius $R_b$ with rates  $k^N_{on}$.  A neurofilament bound  to a nearby microtubule can unbind with rate $k^N_{off}$, or move away and exit $D$ with rate $k^N_{out}$.  We conserved neurofilament number in $D$ by replacing each departing neurofilament with a new entering neurofilament, placed at a distance $R_b$ from a randomly selected microtubule. Neurofilaments and microtubules are long polymers that are aligned in parallel along the long axis of the axon so a new neurofilament can only enter $D$ if a neurofilament was present at that location in the adjacent plane at the preceding time point.  Thus, to prevent the entry of a new neurofilament in a region of $D$ that is lacking other neurofilaments (specifically, this would be encountered when simulating the remixing of neurofilaments and microtubules after segregation), we only permitted the entry of a new neurofilament next to a microtubule that was already within a radius  $R_b$ of another neurofilament.  We did not differentiate the direction of neurofilament movement along the axon because anterograde and retrograde movements have similar contributions to the distribution of neurofilaments in $D$. If a neurofilament is bound to a microtubule, they interact through the following elastic spring forces, 
\beq\label{eqn_force_MN}
\bG^{MN}_{i,j} =  - \bG^{NM}_{j,i} =  
 \kappa^N   d^{MN}_{ij}   \bolde^{MN}_{ij}.  \eeq
Here $\bG^{MN}_{i,j}$ and $\bG^{NM}_{j,i}$ are the forces acting on the $i$-th microtubule by the $j$-th neurofilament and vice versa, and $\kappa^N$ is the spring constant. If $d^{MN}_{ij}$ is bigger than the binding radius $R_b$, then there is no spring force between the neurofilament-microtubule pair.  

We assumed that each neurofilament could engage with only one microtubule at a time.  The rationale for this is as follows.  The on-rate and off-rate for neurofilament binding to microtubules is estimated to be $10^{-2}/\mathrm{s}$ and $6.5\times 10^{-2}/\mathrm{s}$ based on previous experiments.  Thus a neurofilament within the binding radius of a microtubule would spend, on average, $1/(1+6.5)\approx 0.13$ of its time engaged with that microtubule, and the chance for one neurofilament within the binding radius of two microtubules to bind both simultaneously would be $0.13^2 \approx 0.017$. Since in reality no neurofilament would remain within the binding radius of two microtubules at all times, the actual probability is even lower.  Thus the chance for one neurofilament to interact with multiple microtubules simultaneously is negligible and we neglect it in our model. 

\subsubsection*{Mechanism 2: Fast axonal transport of organelles. } 
Like neurofilaments, membraneous organelles are also conveyed anterogradely or retrogradely along microtubule tracks by kinesin and dynein motors.  However, these cargoes tend to spend much less time pausing, resulting in a much faster average rate of transport.  Due to their large size, the movement of organelles can cause significant fluctuations of the microtubule and neurofilament organization by displacing these polymers laterally. These cargoes can bind multiple motors \cite{Gross2007, Mallik2009} and, due to their large size, they can readily interact with multiple microtubules even if those microtubules are not close to each other \cite{Hirokawa1982,Wortman2014}.  As an organelle moves along several microtubules, it can pull them closer together, similar to a ``zipper".  This speculation is supported by {\em in vivo} data that demonstrate organelles being surrounded by multiple microtubules in close proximity \cite{Price1991,Hirokawa1982,Friede1970, Wortman2014}, and {\em in vitro} experiments \cite{Ziebert2009} which show that motors bound to spherical cargoes can pull on multiple microtubules and align them. 

Based on the above considerations, we modeled organelle movement in $D$ in the following way.  Organelles enter $D$ randomly with rate $k^O_{in}$, moving along randomly chosen microtubules, and move persistently until they leave $D$ completely. Thus each organelle is present in $D$ for a time period that equals its length ($2a$) divided by its speed $s^O$, and 
\beq\label{eqn_z}
z_i^O = -a + s^O t,  \qquad   0\leq t \leq 2a/s^O, 
\eeq
where $t$ is the time the organelle has been present in $D$. As the organelle moves from one side of $D$ to the other, its cross-sectional radius $r_i^O$  first increases from 0,  reaches its maximum when it is halfway through, and then decreases to 0, which is given by Eqn \re{eqn_radius_FC}. While present in $D$,  an organelle can bind stochastically to an available microtubule within a binding radius $R_b$ with rate $k_{on}^C$ and unbind with rate $k_{off}^C$. If an organelle and a microtubule are bound, they interact  through the  linear spring force, 
\beq\label{eqn_force_MO}
\bG^{MF}_{i,j} =  - \bG^{FM}_{j,i} =   
\kappa^O   d^{MF}_{ij}   \bolde^{MF}_{ij}.  \eeq
Here $\bG^{MF}_{i,j}$ and $\bG^{FM}_{j,i}$ are the forces acting on the $i$-th microtubule by the $j$-th organelle and vice versa, and $\kappa^O$ is the effective spring constant which represents the action of possibly multiple motors.  

\subsubsection*{Mechanism 3: Volume exclusion. }
 
In addition to the active movement of neurofilaments and organelles and their interactions with microtubules through molecular motors, all the particles in the system interact through forces of volume exclusion. 

Neurofilaments have sidearms which are highly-charged unstructured polypeptide domains. These sidearms project outward from the filament core to form an entropic brush that defines a zone of exclusion around the polymer via long-range repulsive forces \cite{Mukhopadhyay2004,Beck2010a,Stevens2010,Stevens2011,Zhulina2009}, maximizing the space-filling properties of these cytoskeletal elements. Microtubule associated proteins such as tau also have highly charged long polypeptide domains that can have a similar volume-excluding effect \cite{Shahani2002,Brandt2005,Mukhopadhyay2004,Jho2010}.  Based on these biological considerations, we modeled volume exclusion of neurofilaments, microtubules and organelles through the following pairwise repulsions,

\beq\label{eqn_repulsion1}
\bold{R}^{k l}_{i,j} =
\left\{
\begin{aligned}
&  - \eps^{kl}  \Big( L_r\Big/d^{kl}_{ij} - 1\Big) \bolde^{kl}_{ij} && \iif d^{kl}_{ij} \leq   L_r \\
& 0 && \iif d^{kl}_{ij} >  L_r. 
\end{aligned}\right.  
 \eeq
Here $\bold{R}^{k l}_{i,j}$ is the force acting on the $i$-th particle of $k$-type by the $j$-th particle of $l$-type, where 
 $k, l = M, N$ or $O$,  $ 1 \leq i \leq n^{k }$, and $ 1 \leq j \leq n^{l}$. For example,  $\bold{R}_{i,j}^{MN}$ is the force acting on the $i$-th microtubule by the $j$-th neurofilament. Here $L_r$ is the maximum interaction distance;  $\eps^{kl}$ specifies the magnitude of the force; and the negative sign preceding $\eps^{kl}$ indicates that the force is repulsive. We note that this  force goes to infinity as the surfaces of two particles approach each other and remains zero if the distance between two particles is larger than $L_r$. The functional form of the force is similar to those used in \cite{Kumar2002a,Kumar2002} for neurofilament repulsions and matches recent experimental data \cite{Srinivasan2014}. 
 
To keep all the particles inside the domain, we modeled volume exclusion of the particles with the domain boundary in a similar way.  The force acting on the $i$-th particle of $k$-type by the axonal membrane is given by
\beq\label{eqn_repulsion2}
\bold{R}^{kB}_{i} =  
\left\{
\begin{aligned}
& - \eps^{kB}  \Big( L_r\Big/d^{kB}_{i} - 1\Big) \bolde^{kB}_{i}   && \iif d^{kB}_{i} \leq   L_r \\
& 0 && \iif d^{kB}_{i} >  L_r.
\end{aligned}\right.
\eeq
Here the index $B$ stands for ``boundary", 
$d^{kB}_{i} = R_0 - | \bx_i^k| - r^k$, and $\bolde^{kB}_{i}$ is the unit vector pointing from the center of the domain to $\bx_i^k$.

Microtubules, neurofilaments, and organelles can also interact with each other hydrodynamically through the axoplasm. Organelle movement can cause significant flow of the axoplasm near their surfaces and displace nearby microtubules and neurofilaments. As an organelle pushes into $D$, its radius increases and it pushes nearby fluid and particles away from itself;  as it moves away from $D$, instead of leaving void behind it, it creates negative pressure which draws the axoplasm to flow back and fill the space. The hydrodynamic effect due to the movement of microtubules and neurofilaments is presumable smaller given their constant and smaller size in cross-section. In this model, we do not model the hydrodynamic interactions among these particles explicitly, but include this effect by adjusting the force prefactors associated with organelles. Specifically, when an organelle push into the domain, we double $\eps^{kO}$ and $\eps^{Ok}$  to take into account the contribution of the fluid flow it creates.

\subsubsection*{Model equations. }
The movements of microtubules, neurofilaments, and organelles in axons are viscous-dominated and thus inertia can be neglected.  Under this simplification, we have the following system of stochastic differential equations
\beq\label{eqn_main}
 \md\bx^k_i = \bF_i^k  /\mu^k \, \md \rm{t} + \sigma_k \, \md \bW_i^k, \quad 1 \leq i  \leq  n^k,  \; k = M, N, F.
\eeq
Here $\bF_i^k$  is the sum of all applied forces on that particle specified in \re{eqn_force_MN}, \re{eqn_force_MO}, \re{eqn_repulsion1} and \re{eqn_repulsion2}. For example,   
$
\bF_i^N = \sum_j \bG_{i,j}^{NM} + \sum_j  \mathbf{R}_{i,j}^{NM}  + \sum_{j, \, j\neq i} \mathbf{R}_{i,j}^{NN}  +  \sum_j \mathbf{R}_{i,j}^{NO} + \mathbf{R}_i^{NB}. 
$
The constant  $\mu^k$ denotes the drag coefficient of the $k$-type particle.  Finally $\bW_i^k$  are independent 2D Wiener processes modeling the random motion of these particles,  and the amplitude  $\sigma_k$ is given by $\sigma_k = \sqrt{2D_k}$, where  $D_k$ is the diffusion coefficient of the particle calculated by the Einstein relation. 

\smallskip

\subsection*{Parameter estimation and simulation algorithm}

The parameters used in our model are physical, and thus they are all measurable. Most of them have already been measured or there exist experiments that can be used to estimate them.  Table \ref{table_params} summarizes all the parameter values, and the detailed methods to obtain these parameters are given in the SI.  The units of these parameters reflect the time scales for the molecular processes integrated into the model, which are seconds or fractions of a second. 

\begin{table}[!ht]
\begin{adjustwidth}{-0.5in}{0in} 
\caption{ {\bf Model Parameter Values} }
\medskip
\begin{tabular}{cp{0.42\textwidth}p{0.24\textwidth}p{0.18\textwidth}}
\hline 
Parameter & Description & Values& Notes and Refs  \\  \hline  
\tskip 
$r^N$ & Radius of neurofilament backbone &  $5$ nm & \cite{Janmey2003,Fuchs1998,Brown2013} \tskip  \\ 
$r^M$  & Radius of microtubule backbone & $12.5$ nm & \cite{Brown2013} \tskip \\ 
$R_b$ & capturing radius for microtubule-cargoes active binding & $80$ nm &   \cite{Hirokawa1989} \tskip \\  
$k_{on}^N$ & rate for neurofilament binding & $1.0\times 10^{-2} / \mathrm{s}$ &  \cite{Li2012}, E.E.  \tskip \\
$k_{off}^N$ & rate for neurofilament unbinding &  $6.5\times 10^{-2} / \mathrm{s}$  &  \cite{Li2012} \tskip \\
$k_{out}^{N}$ & rate for neurofilament departure &  $0.1\, \mathrm{s}^{-1}$  & E.E \tskip \\
$k_{on}^O$ & rate for organelle binding & $2 / \mathrm{s}$ &  \cite{Kunwar2008,Erickson2011}  \tskip \\
$k_{off}^O$ & rate for organelle unbinding &  $2 / \mathrm{s}$  & \cite{Kunwar2008} \tskip \\
$k_{in}^O$ &  rate for organelle passage   & $0.105/\mathrm{s}$ & \cite{Smith1980,Papasozomenos1981}, E.E. \tskip \\ 
$s^O$ & speed of organelle movement along microtubules & $1\,\micron/\rms$ & \cite{Smith1993} \tskip \\ 
$L_r$ & characteristic repulsion distance  & 121.2 nm &  \cite{Kumar2002}, E.E.   \tskip \\
$\eps_r$ & repulsion scale ( = $\eps^{NN}$) & 0.5 pN   &  E.E. \tskip \\
$\kappa^N $ & effective spring constant for microtubule-neurofilament binding  & $0.18 \pN/\mbox{nm}$  &   \cite{Coppin1995}, E.E.  \tskip \\
$\kappa^O $ & effective spring constant for organelle-microtubule binding  & $0.9 \pN/\mbox{nm}$ & \cite{Coppin1995}, E.E.   \tskip \\
$\mu^N$  & drag coefficient of neurofilaments & 73.5  $\mathrm{pN} \cdot \mathrm{s} / \micron$   &  E.E \tskip \\
$\mu^M$  & drag coefficient of microtubules &  512 $\mathrm{pN} \cdot \mathrm{s} / \micron$  &   E.E \tskip \\
$\mu^O$  & drag coefficient of organelles &  40.3$\mathrm{pN} \cdot \mathrm{s} / \micron$ &  E.E \tskip \\
$D_N$  & diffusion coefficient of neurofilaments &  $5.59 \times 10^{-5} \mu m^2/ \mathrm{s} $ & E.E.  \tskip \\
$D_M$  & diffusion coefficient of microtubules &  $8.02 \times 10^{-6} \mu m^2/ \mathrm{s} $ &  E.E. \tskip \\
$D_F$  & diffusion coefficient of organelles &   $1.02 \times 10^{-4} \mu m^2/ \mathrm{s} $  &  E.E.  \tskip \\
\hline
\end{tabular}
\begin{flushleft}
{E.E.: estimated from experiments; see SI for detailed information. }
\end{flushleft}
\label{table_params}
\end{adjustwidth}
\end{table}%

To solve the model numerically, we treated the binding and unbinding, arrival and departure of cargoes explicitly at discrete time steps, and integrated the model system (\ref{eqn_main}) using the explicit Euler's method. Because $\sigma_k$, $k = M, N, C$ are constant in time, the numerical integrator has  strong order 1.0  \cite{KloedenPlatenBook}. We chose a time step $h$ much smaller than all the time scales involved in Mechanisms 1-3.   For the simulations of segregation and remixing over hours to a day, we used $h = 1/50 \sec$ if there was no organelle in $D$,  and $h = 1/1600 \sec$ otherwise in order to deal with the  stiffness of the equations introduced by the pushing of organelles when they move into $D$. The detailed simulation algorithm is included in the SI. The computational tool is written in C++.  

\section*{Results}

\subsection*{The organization of neurofilaments in normal axons. }
\label{sec_3.1}

Morphometric studies suggest that neurofilaments are spaced randomly in axonal cross-sections when packed at low densities, but as the density increases they start to experience the volume-exclusionary repulsive forces of their neighbors and assume a less random distribution characterized by a more even neurofilament spacing \cite{Price1988,Hsieh1994,Kumar2002}. In this section we demonstrate that the neurofilament distribution generated using our model agrees well with these experimental data. 

Different methods have been used to characterize neurofilament distribution in axonal cross-sections.  Kumar et al \cite{Kumar2002} used the radial distribution function (RDF) (also known as the pairwise correlation function). The RDF, denoted as  $g(r)$ describes how density varies as a function of distance from a reference particle. For particles that move randomly and completely independently, $g(r)$ is a constant value of 1;  while for crystalline structures $g(r)$ forms peaks at precisely defined intervals. For neurofilaments in axons the shape of $g(r)$ typically lies between these two extremes, increasing sharply from 0 and forming a peak around $30 - 50 \, \rm{nm}$ \cite{Kumar2002}. Another method used often is to calculate the occupancy probability distribution (OPD), which is the distribution for the number of particles within an observation window of a specified shape and size \cite{Price1988,Hsieh1994,Kumar2002}. For neurofilaments, the OPD can be approximated by Guassian  \cite{Hsieh1994,Kumar2002}. 

\begin{figure}[h]
\begin{center} 
\includegraphics{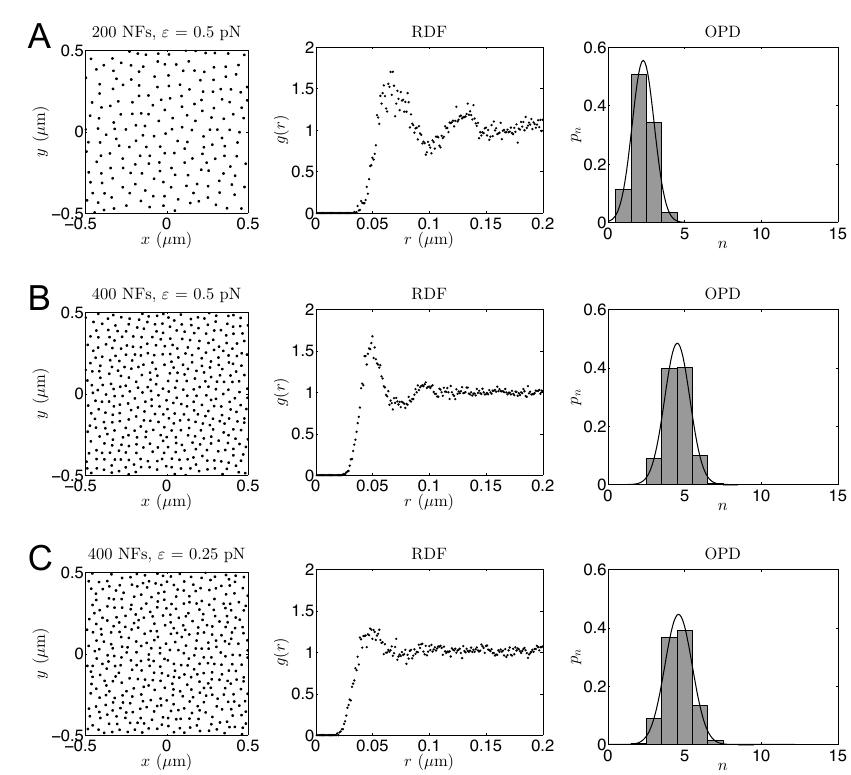}  
 \end{center}
\caption{{ Simulated neurofilament distributions with different densities and repulsion magnitudes.} Left: snapshots of neurofilament positions after randomizing for 25 sec. Middle: the radial distribution functions (RDF,  $g(r)$). Right: the bars are histograms of the occupancy probability distribution (OPD, $p_n$) using randomly chosen circular windows with a radius 60 nm, and the black curves are their Gaussian fits. Middle and right plots represent averages over 50 time frames between $ t = 25 \, \rm{sec}$ and $ t = 30 \, \rm{sec}$. In all cases, $\eps$ is short for $\eps^{NN}$. (A) $n^F =  200$, $\eps^{NN} = 0.5$ pN; (B) $n^F =  400$, $\eps^{NN} = 0.5$ pN; (C) $n^F =  400$, $\eps^{NN} = 0.25$ pN. Other parameters are the same as specified in Table \ref{table_params}. }\label{fig_neurofilament_distribution_1}
\end{figure}

In previous experimental studies, the RDF and OPD of neurofilaments were calculated in selected regions of axonal cross-sections with almost no microtubules and organelles. To mimic such conditions, we performed simulations with exclusively neurofilaments, i.e., $n^M = n^O = 0$, and thus the only acting mechanisms are the pairwise repulsions and the Brownian motion of neurofilaments.  We used a square domain with side length $1 \micron$, and to minimize the effect of the boundary we used periodic boundary conditions. Under such conditions, the system \re{eqn_main} reduces to 
\beqn
 \md\bx^N_i = \sum_{j, j\neq i}\bold{R}_{i,j}^{NN}  /\mu^N \, \md \rm{t} + \sigma_N \, \md \bW_i^N, \qquad 1 \leq i  \leq  n^N. 
\eeqn
We initially put neurofilaments on a hexagon lattice inside the domain, and then "randomized" the distribution by simulating the model for sufficient time to observe no further change in the OPD or RDF.  To solve the model, we used the explicit Euler's method with a time step $ h =  1/200 \sec$.  
  
We first investigated how the neurofilament distribution depends on its density. We took $\eps^{NN}  = 0.5 \mathrm{pN}$ and used increasing neurofilament densities of 200  and 400 per $\micron^2$ (Fig. \ref{fig_neurofilament_distribution_1}, Rows A and B). For each case, the left panel is a plot of the coordinates of the neurofilaments after randomizing for 25 sec;  the middle panel is a plot of the RDF which represent averages over 50 time frames between 25 sec and 30 sec; and the right panel is a plot of the averaged OPD and its Gaussian fit. The methods that we used to calculate the RDF and OPD are the same as in \cite{Kumar2002} and described in the supporting information (SI). These plots show that as the neurofilament density becomes higher, the separation of the peaks of the RDF becomes smaller,  and the average and variance of the OPD becomes larger as the neurofilament density becomes larger.  General features of these plots are in tight agreement with experimental data presented in \cite{Price1988,Hsieh1994,Kumar2002}.

The magnitude of the repulsion between two neurofilaments depends on the charges of their sidearms. As the phosphorylation level of their sidearms becomes higher,  their mutual repulsion becomes larger. We next investigated how the neurofilament distribution depends on the effect of sidearm phosphorylation by fixing the neurofilament density  and varying $\eps^{NN}$. We took the neurofilament density to be 400 per $\micron^2$, and  $\eps^{NN}$ to be $0.25 \pN$ and  $0.5 \pN$.  Fig. \ref{fig_neurofilament_distribution_1}B and C shows that as  $\eps^{NN}$ becomes larger, the locations of neurofilaments become more regular, the peaks of the RDF are better defined, and the variance of the OPD becomes smaller. 

\subsection*{Impairment of neurofilament transport leads to microtubule-neurofilament segregation. \, }

To investigate the mechanism of microtubule-neurofilament segregation in axons, we compared our simulations to experimental data obtained for IDPN in laboratory animals.  We focused on IDPN because there is published data on both the rate and reversibility of the segregation.  When IDPN is administered transiently by local injection into peripheral nerves, segregation appears within 2-6 hours and then disappears within 24 hours \cite{Griffin1983,Papasozomenos1981,Nagele1988a}.

\begin{figure}[h]
\includegraphics[width=0.9\textwidth]{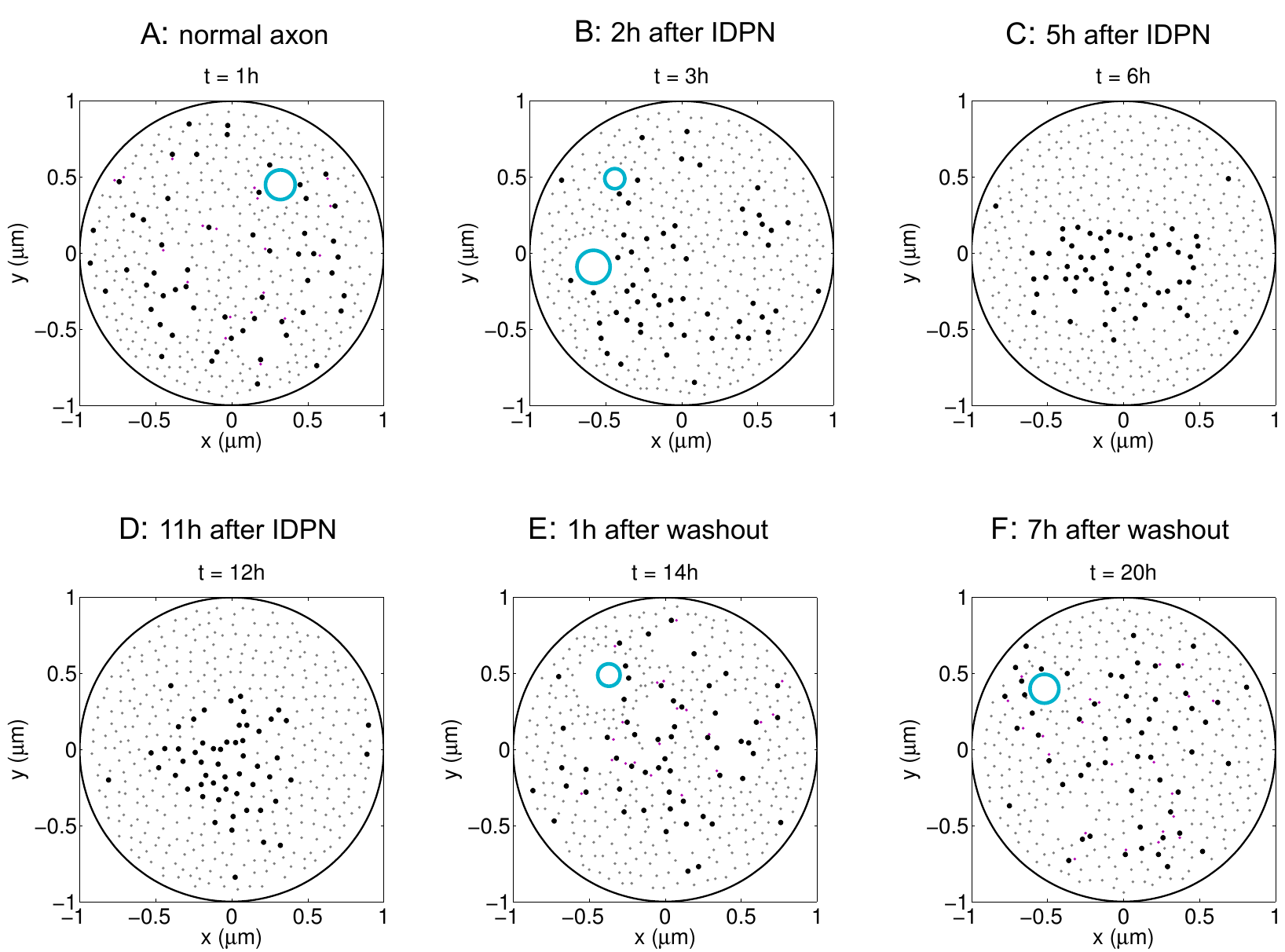}
\caption{ { Reversible segregation of microtubules and neurofilaments.} neurofilament transport is blocked starting at $t=1 \, \rm{ h}$ and restored at $t = 13 \, \rm{ h}$. (A-F) Snapshots of the positions of microtubules, neurofilaments and organelles at $t = 1 \hour$, $3 \hour$, $6 \hour$, $12 \hour$, $14 \hour$, $20 \hour$. All panels are from a single realization of the model.  (A) microtubules and neurofilaments form a mixture under normal conditions. (B-D) blockage of neurofilament transport leads to gradual segregation of microtubules and neurofilaments.  (E,F) restoration of neurofilament transport causes remixing of microtubules and neurofilaments. Small grey dots are free neurofilaments; small purple dots are neurofilaments engaged with microtubules; large black dots are microtubules; large blue circles are organelles. Parameters used: $n^M = 56, n^N = 361$.  Neurofilament on-rate $k_{on}^N$ equals  0 between $t = 1\hour$ and $13\hour$.     All other parameters are the same as in Table \ref{table_params}. }\label{fig_seg}
\end{figure}

Since neurofilament accumulation and axonal swelling occur on a much slower time course than the segregation, they can be ignored for the purposes of our current analysis.  Therefore, for simplicity, we took $D$ to be a disk with fixed radius $R_0 = 1 \micron$, and set the total number of neurofilaments $n^N$ to be constant. Specifically, if a neurofilament that was engaged with a microtubule left $D$, then it was replaced by a new neurofilament that entered $D$ by association with a new randomly chosen microtubule. The total number of microtubules and neurofilaments in the domain were determined based on the experimentally determined densities of $18 /\micron^2$ and $115 /\micron^2$, respectively \cite{Papasozomenos1981}.  We thus calculated $n^M$ by the formula $n^M = \mbox{floor}(18 \pi R_0^2) = 56$ and similarly we obtained $n^N = 356$. Here the function $\mbox{floor}(u)$ is the largest integer that is smaller than $u$. We considered organelles with $b = 140 \, \mbox{nm} $ and $a/b = 10$ (Fig. \ref{fig_geometry}C). 

We started the simulations by including axonal transport of both neurofilaments and organelles, mimicking the conditions of normal axons.  To distribute the neurofilaments and microtubules randomly without overlap, we first placed them on a hexagon lattice in $D$ with no organelles, and then introduced volume exclusion and Brownian motion for enough time to randomize their positions. Starting from this initial condition, we then turned on the movement of both neurofilaments and organelles. Fig. \ref{fig_seg}A is a snapshot of the simulated distribution of microtubules, neurofilaments, and organelles in a normal axon. The small grey dots are neurofilaments that are not engaged with microtubules, the small purple dots are neurofilaments that are engaged with microtubules, the large black dots are microtubules, and the large cyan circle is an organelle pushing into the cross-sectional domain. Note that a small fraction of the neurofilaments are bound to microtubules and moving along microtubules, that one microtubule can transport multiple cargoes (neurofilaments or organelles), and that one organelle can engage with multiple microtubules simultaneously.  

We then blocked neurofilament transport selectively by resetting the binding rate of neurofilaments to microtubules, $k_{on}^N$, to be 0 at $t = 1\mathrm{h}$. This disengaged neurofilaments from their microtubule tracks and thus blocked their movement so that none could enter or leave $D$. Meanwhile, the transport of organelles was not affected: they continued to grab microtubules stochastically, pulling them together. This ``zippering'' effect caused the microtubules to gradually cluster (\ref{fig_seg}B). By 6 hours, almost all the microtubules had migrated to the center of $D$ and formed a single island surrounded by neurofilaments (\ref{fig_seg}C, D).  The central microtubule cluster contained organelles but relatively few neurofilaments whereas the peripheral zone of neurofilaments contained relatively few microtubules or organelles.  This segregation pattern is strikingly similar to that observed in experiment and in disease, and the rate of segregation is comparable to that observed experimentally for local injection of IDPN into peripheral nerves of laboratory animals \cite{Griffin1983,Papasozomenos1981}. 

After observing segregation, we restored neurofilament transport by resetting $k_{on}^N$ to its original value at $t=13\, \mathrm{h}$. This immediately allowed neurofilaments on the periphery of the microtubule core within a distance $R_b$ of a microtubule to bind to that microtubule stochastically and then either unbind or exit $D$ after a short while, as dictated by their stop-and-go transport behavior.  As explained in the Methods, each neurofilament that exited $D$ was replaced with a new neurofilament seeded adjacent to a randomly selected microtubule, but only if that microtubule was within a distance of $R_b$ from another neurofilament already in that plane.  Over time this resulted in a gradual infiltration of neurofilaments into the microtubule cluster in a centripetal manner (i.e. from the outside edges progressing inward), leading to a gradual dispersal of the microtubules (\ref{fig_seg}E) and a return to their normal interspersed organization (\ref{fig_seg}F). These results agree tightly with previous experimental findings.

\begin{figure}[h]
\centering
\includegraphics{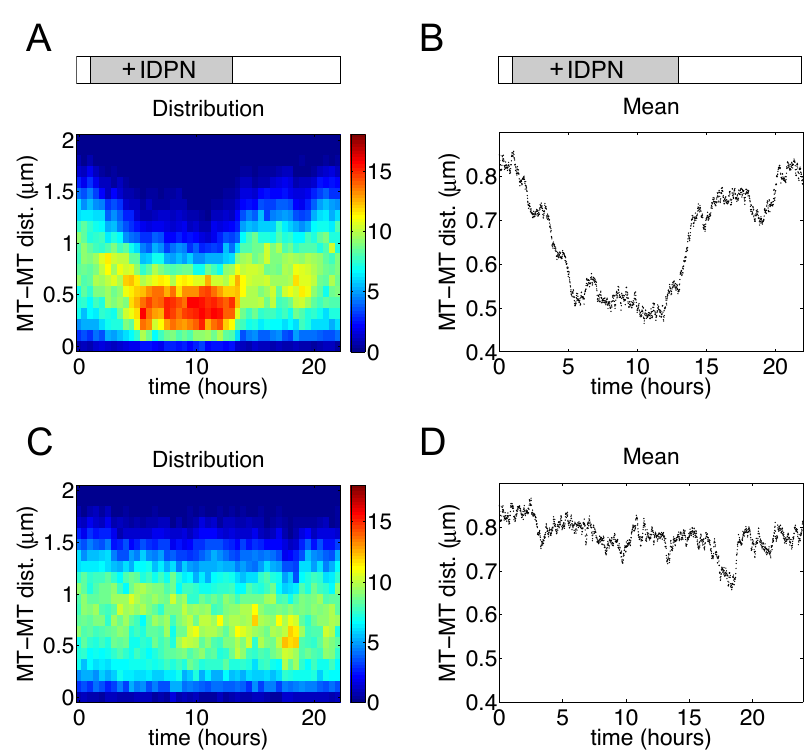}
\caption{ { Statistics of the pairwise distances between microtubules (PDMT). }  (A, B) IDPN treatment started at t=1 hour and stopped at t=13 hours. (C, D) control. (A, C) distribution of the PDMT; data plotted for every 20 min. The pseudo color key represents the number of microtubule pairs.  (B, D) mean of the PDMT; data plotted for every min.  Parameters used are the same as in Fig. \ref{fig_seg}.  }\label{fig_MTdistance}
\end{figure}

To characterize the reversible segregation of microtubules and neurofilaments, we plotted the distribution and the mean of the pairwise distance between two microtubules (PDMT) as a function of time. Figs. \ref{fig_MTdistance}A and  \ref{fig_MTdistance}B are plots calculated from the simulation shown in Fig. \ref{fig_seg}, which demonstrate a significant progressive decrease of the PDMT upon elimination of neurofilament transport  ($t=1\mathrm{h}$) and subsequent increase upon restoration of neurofilament transport ($t=13\mathrm{h}$). Figs. \ref{fig_MTdistance}C and  \ref{fig_MTdistance}D are plots for a normal axon for comparison. We see that under normal conditions, because microtubules and neurofilaments are interspersed, the distribution of PDMT is broad and the mean of it is about $0.8R_0$; and as microtubules and neurofilaments segregate from each other, the distribution becomes more compact and the mean of the PDMT decreases by almost 40\%.

Another way to incorporate blockage of neurofilament transport is to increase the off-rate of neurofilaments $k_{off}^N$. We performed simulations with $k_{off}^N$ 100 times larger, and obtained similar results as in Fig. \ref{fig_seg}.

\subsection*{Microtubule zippering by moving organelles is the causal mechanism for segregation. \,}

In the above section we have shown that in the absence of neurofilament transport, organelle transport leads to microtubule-neurofilament segregation. As we noted earlier,  organelles can interact with multiple microtubules simultaneously and thus pull or zip nearby microtubules closer together.  We next investigated the importance of this zippering mechanism for the segregation of microtubules and neurofilaments. To do this, we introduced a maximum number of microtubules that a single organelle can interact with simultaneously, denoted by $m_{\max}$, and investigated how the PDMT depends on $m_{\rm max}$ in the absence of neurofilament transport. 

\begin{figure}[h]
 \begin{center} 
 \includegraphics[width=0.4\textwidth]{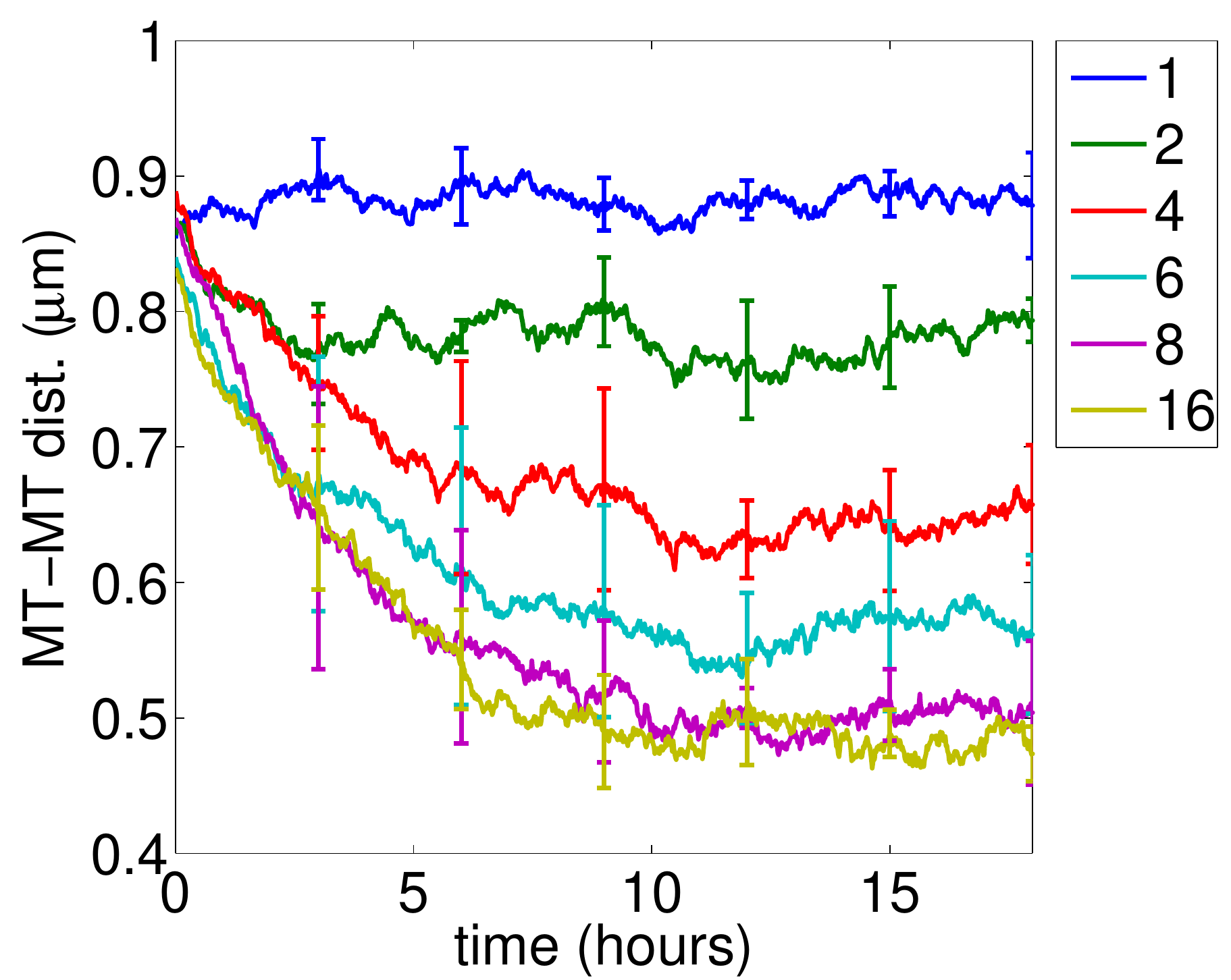} 
\end{center}
\caption{ { The effect of microtubule zippering by moving organelles.} The mean of PDMT is plotted over time. The maximum number of microtubules  that a single organelle can interact with simultaneously ($m_{max}$) is set to be 1, 2, 4, 8, and 16 for the blue, green, red, cyan, purple, and yellow curves respectively. Each curve represents the average over 5 realizations of the model and the error bars are the standard deviation.  All other parameters are the same as in Fig. \ref{fig_seg}.  }\label{fig_seg_mmax}
\end{figure}

Fig. \ref{fig_seg_mmax} plots the mean of PDMT as a function of time given different values of $m_{\rm max}$. Each curve is averaged over five realizations with unpredictable seeds, and the error bars indicate the standard deviations over the realizations. If each organelle is only allowed to bind to one or two microtubules, i.e., $m_{\rm max} = 1$ or $2$, then microtubules and neurofilaments remain mixed over time and segregation does not occur at all (blue and green).  Indeed, for $m_{\rm max} = 1$, the mean of PDMT is slightly larger than that for a normal axon shown in Fig. \ref{fig_MTdistance}D. This is because organelles stir microtubules and neurofilaments and separate microtubules apart.  As $m_{\rm max}$ increases, the PDMT curve decreases faster and the time needed to reach complete segregation decreases. Scatter plots of microtubules and neurofilaments (not shown here) show that for $m_{\max} = 4$, partial but significant segregation was observed by 18 hours in all five realizations; for $m_{\max} = 6$,  complete segregation was observed  by 18 hours in four out of five realizations; for $m_{\max} = $ 8 or 16, complete segregation was observed in all realizations within 10 hours. These results suggest that microtubule zippering by moving organelles is the causal mechanism for the segregation of microtubules and neurofilaments in the absence of neurofilament transport. 

\subsection*{Dependence on the size and the flux rate of organelles. \, }

\begin{figure}[h!]
\begin{center} 
 \includegraphics[width=0.38\textwidth]{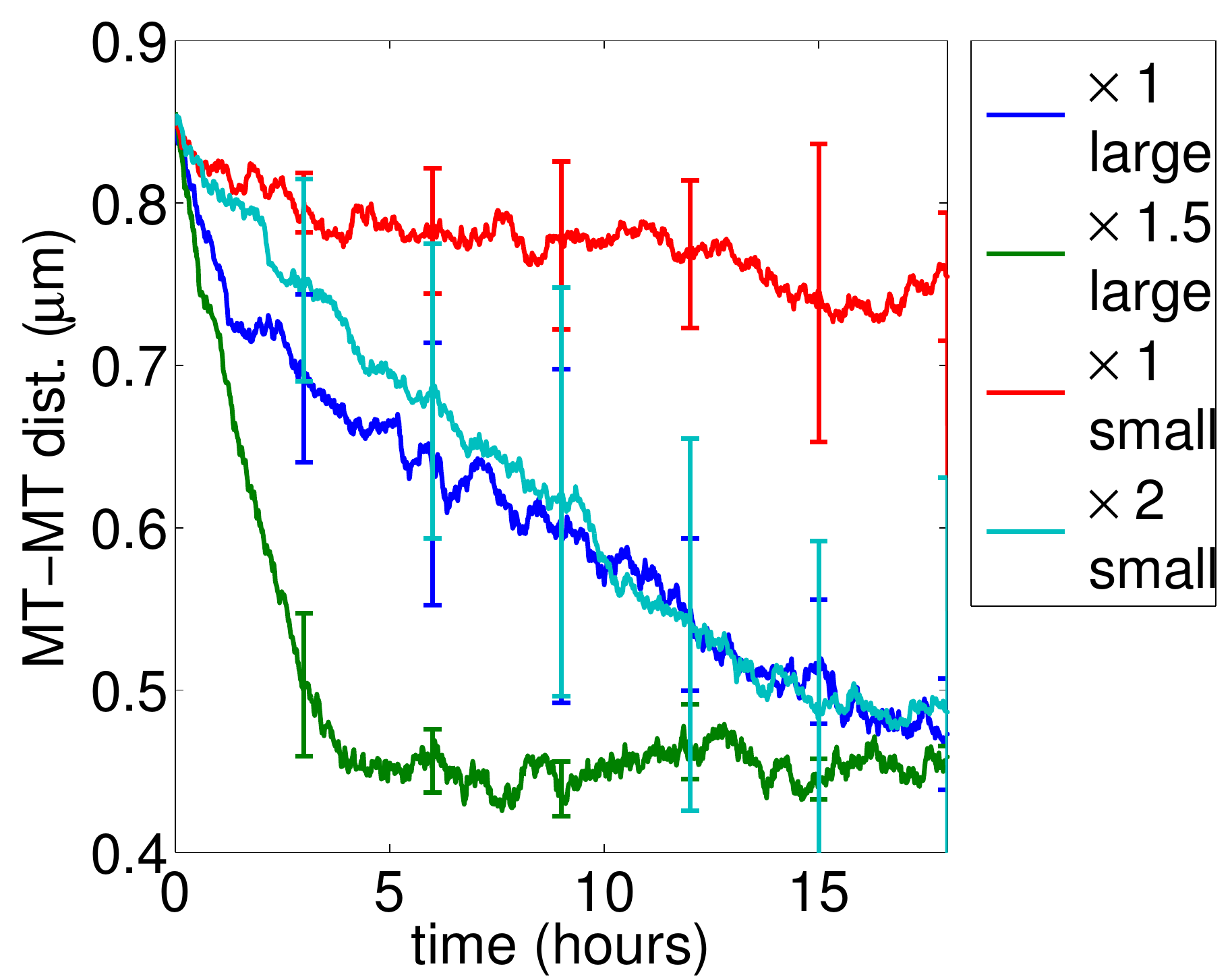} 
\end{center}
\caption{ { The segregation process depends on organelle size and  flux rate.} The mean PDMT is plotted over time. Organelle max radius $b$:  $140\, \mathrm{nm}$ for blue and green curves (same as Fig. \ref{fig_seg});  $b = 70\, \mathrm{nm}$ for red and cyan curves.  Organelle flux rate $k_{in}^O$:  same as Fig. \ref{fig_seg} (x1; blue and red), 1.5-fold greater (x1.5; green) and 2-fold greater (x2; cyan).  Each curve represents an average of 5 realizations of the model and the error bars are the standard deviation.  All other parameters are the same as in Fig. \ref{fig_seg}.  }\label{fig_size}
\end{figure}

We next investigated the dependence of the segregation on the size and the flux rate of the organelles by simulating the model with different sizes of organelles, $b = 140 \, \mathrm{nm}$ or $b = 70 \, \mathrm{nm}$, and different flux rates $k_{in}^O$. Fig. \ref{fig_size} plots the mean PDMT over time for four situations: $b = 140 \, \mathrm{nm}$ and $k_{in}^O = 0.105 /\mathrm{s}$ (shown in blue), which is the same as in Fig. \ref{fig_seg}; $b = 140 \, \mathrm{nm}$ and $k_{in}^O = 0.1575 /\mathrm{s}$ (shown in green); $b = 70 \, \mathrm{nm}$ and $k_{in}^O = 0.105 /\mathrm{s}$ (shown in red); and $b = 70 \, \mathrm{nm}$ and $k_{in}^O = 0.21 /\mathrm{s}$ (shown in cyan). These results suggest that (1) for organelles of the same size, the more frequently they move through $D$, the faster the segregation occurs; (2) given the same flux rate across $D$, larger organelles are more capable of clustering microtubules and segregating them from neurofilaments than small organelles, and this is because on average larger organelles can interact with more microtubules simultaneously.

\begin{figure}[h!]
\begin{center} 
 \includegraphics{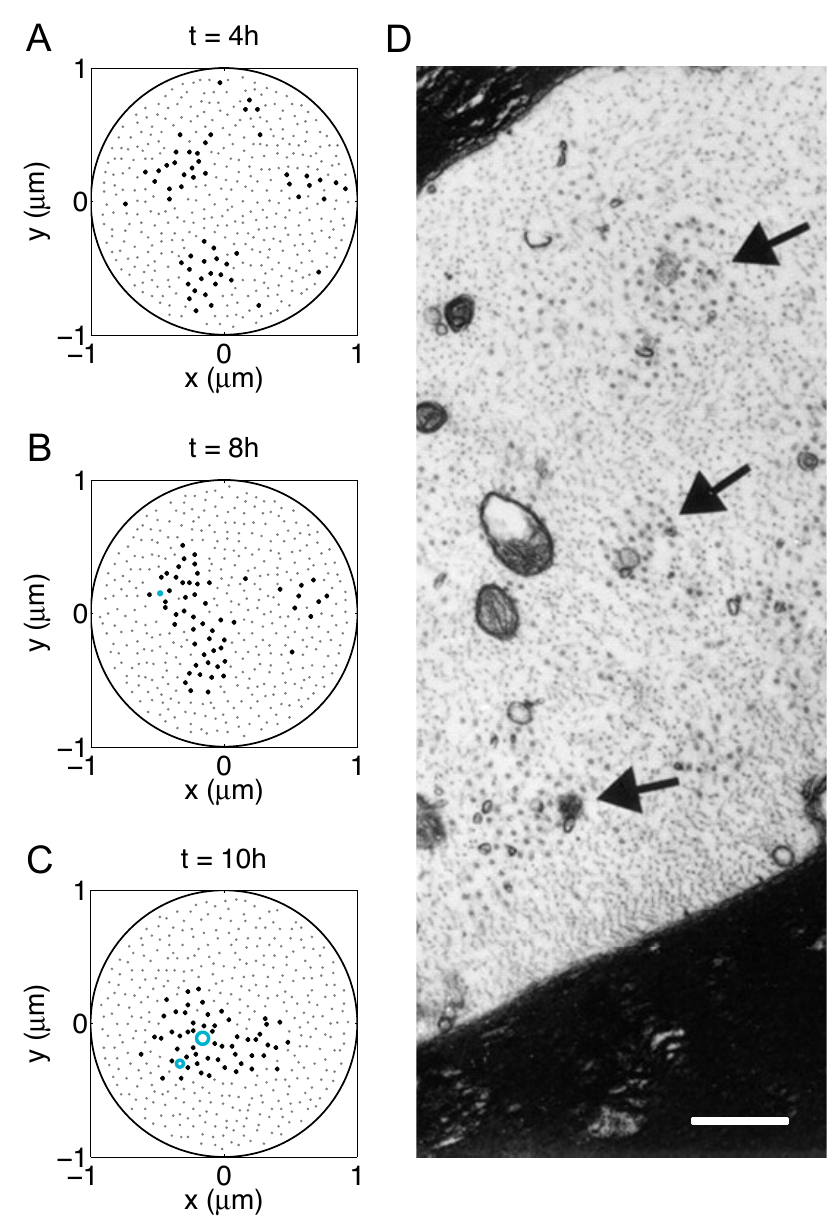}  
 \end{center}
\caption{ { Intermediate states of the segregation process}. (A-C) snapshots of the segregation process in a single realization. (A) microtubules form three clusters by $t=4\mathrm{h}$. (B) these clusters remain separated for several hours until two of them merge around $t = 8 \mathrm{h}$. (C) finally all microtubules form  a single big cluster near the center of the domain. The dimension of the organelles: $b = 70 \mathrm{nm}$, $a/b = 10$. The flux rate of the organelles: 0.21/s.  All other parameters  are the same as in Fig. \ref{fig_seg}. (D) multiple microtubule clusters observed in experiments with IDPN. Big dots: microtubules; circular objects: organelles. Black arrows point to microtubule islands. Adapted from Fig. 9A of \cite{Zhu1998}. The scale bar is 0.4 $\micron$ }\label{fig_intermediate}
\end{figure}

Interestingly, simulations of the model demonstrate that during the segregation process microtubules frequently form smaller clusters first, then these small clusters gradually merge with each other to finally form a single large cluster near the center of the domain. These intermediate states were more apparent in simulations with small organelles, presumably because the rate at which the smaller clusters merge  is slower under this condition. Figs. \ref{fig_intermediate}A-C are snapshots of these intermediate states captured in a single realization (corresponds to the cyan curve in Fig. \ref{fig_size}).  A similar pattern of isolated clusters of microtubules has also been reported by Zhu et al \cite{Zhu1998} (Fig. \ref{fig_intermediate}D) (see Discussion).

\subsection*{Partial blockage of neurofilament transport: dosage effect. \,}

\begin{figure}[ht]
\begin{center} 
 \includegraphics[width=0.4\textwidth]{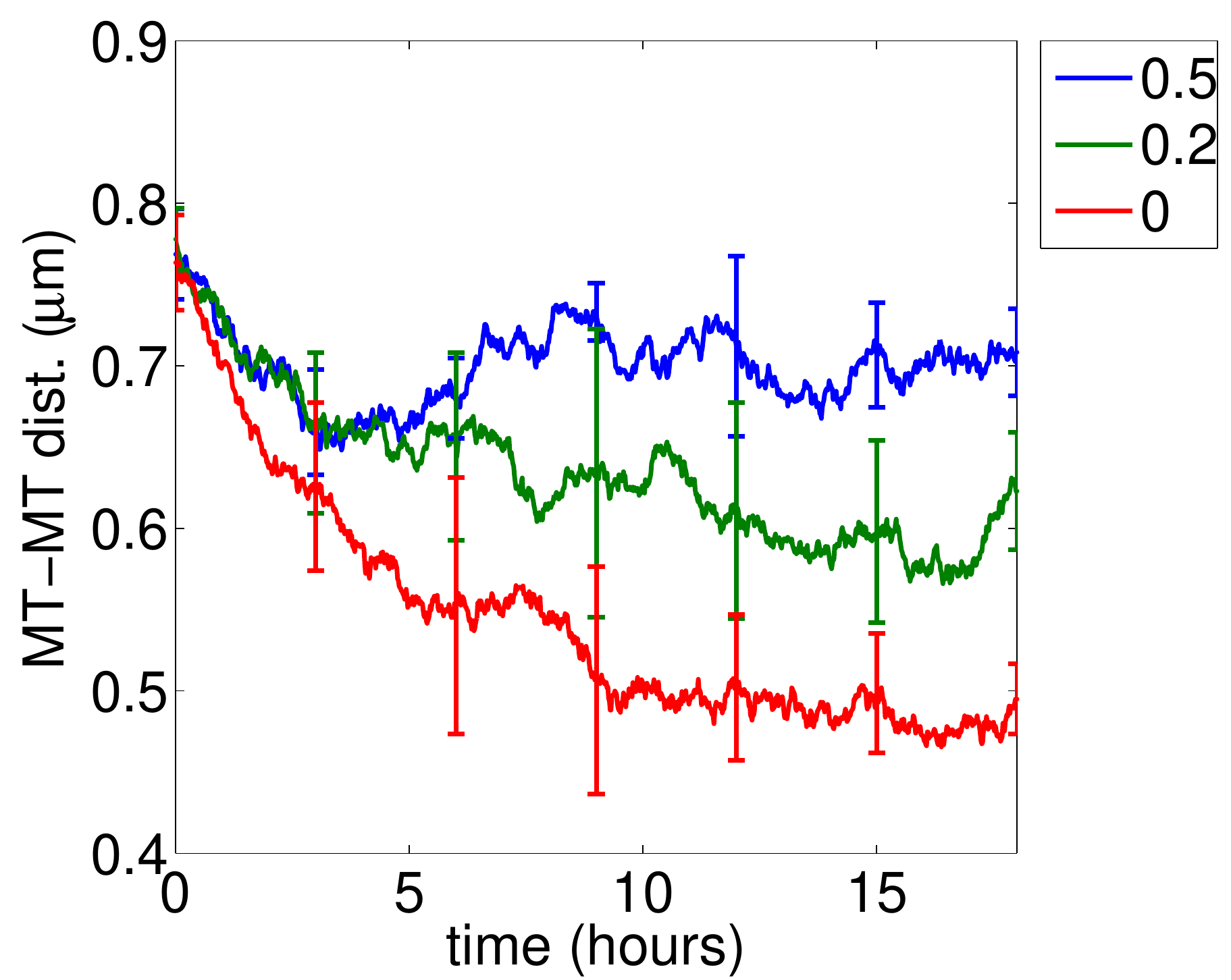} 
 \end{center}
\caption{ { Dependence of  microtubule-neurofilament segregation on $k_{on}^N$. } Each curve plots the mean of the PDMT over time averaged over five realizations, and the error bars are the standard deviations.  The rate $k_{on}^N$ is reduced to  $50\%$ (blue), $20\%$ (green), and $0\%$ (red) of the value in Fig. \ref{fig_seg} at $t = 1\mathrm{h}$. All other parameters  are the same as in Fig. \ref{fig_seg}. }\label{fig_seg_dosage}
\end{figure}

We finally investigated the cross-sectional distribution of microtubules and neurofilaments when neurofilament transport is partially blocked.  In the case of segregation induced by IDPN, this might be considered equivalent to varying the IDPN concentration. To do this we reduced $k_{on}^N$ by different extents at $t = 0\mathrm{h}$. Fig. \ref{fig_seg_dosage} plots the mean of PDMT over time for $k_{on}^N$ equals 0.5, 0.2 and 0 times of its original value. Each curve was obtained by averaging over 5 realizations with unpredictable seeds, and the error bars indicate the standard deviations about the mean. The data indicate that when $k_{on}^N$ is small enough, there is insufficient neurofilament transport to counteract the organelle-dependent microtubule clustering, and segregation is observed.  However, as $k_{on}^N$ becomes larger, the rate of microtubule clustering becomes slower and the resulting clusters become less compact, reflecting less efficient segregation. Increasing $k_{off}^N$ has a similar effect: as $k_{off}^N$ becomes larger, the rate of microtubule clustering becomes faster and the clusters become more compact (not shown). 
 Thus the rate and extent of microtubule-neurofilament segregation is dependent on the extent of inhibition of neurofilament transport.

\section*{Discussion}

\textit{Summary of our model.}  We developed a novel stochastic multiscale model for the cross-sectional distribution of microtubules and neurofilaments in axons. The model describes microtubules, neurofilaments, and membranous organelles as interacting particles in an axonal cross-section. It incorporates detailed descriptions of key molecular processes hat occur within seconds, including the axonal transport of neurofilaments and membranous organelles through this plane, as well as volume exclusion and Brownian motion of all the particles, and addresses the segregation phenomena that occur on a time scale of hours to days. The positions of the particles in the plane are governed by a system of stochastic differential equations.   

Mathematical models of the axonal transport of neurofilaments and organelles have been developed previously to describe the longitudinal distribution of cargoes along axons
\cite {Blum1989,Blum1985,Brown2005,Craciun2005,Jung2009,Li2014,Popovic2011}. However, those models were in 1D and did not consider the spatial arrangement and mechanical interactions of the cargoes and tracks in the radial dimension which are essential in understanding the segregation of microtubules and neurofilaments as well as the subsequent axonal swelling in neurological diseases.  In our model, we describe in detail the dynamic interactions of neurofilaments, organelles, and nearby microtubules through molecular motors and volume exclusion in cross-section. Simulations of the model are in tight agreement with experimental data and generated a number of predictions that can be tested experimentally.  

\textit{Neurofilament and membranous organelle transports are competing processes.}  Simulations of the model demonstrate that if we block neurofilament transport selectively by preventing neurofilament binding to microtubules, while allowing organelle movement to continue, then the moving organelles tend to zipper nearby microtubules together so that they gradually segregate from the neurofilaments.   The microtubule zippering action of the membranous organelles arises because we allow multiple motors to engage with a single organelle, which is consistent with experimental data and theoretical considerations \cite{Gross2007,Shubeita2008,Mallik2009,Erickson2011}.  Restoration of neurofilament transport in the model allows the neurofilaments and microtubules to remix until their spatial distribution returns to normal. This suggests that  neurofilament transport and organelle transport are competing processes in determining the cross-sectional distribution of microtubules: neurofilament transport can insert neurofilaments between adjacent microtubules, pushing those microtubules apart, while organelle transport can pull microtubules together when they move along multiple microtubules simultaneously, similar to a zipper. In normal axons, a dynamic balance between these two processes leads to the interspersed distribution of microtubules and neurofilaments, while in the absence of neurofilament transport, the microtubule zippering effect of organelle transport causes  microtubules and neurofilaments to segregate.   Thus our model predicts that the microtubule-neurofilament segregation that is observed in axons in neurotoxic and neurodegenerative diseases is a simple emergent property of the motile properties of membranous organelles that is triggered by selective impairment of neurofilament transport.  An important and experimentally testable prediction of this study is that segregation is dependent on organelle movement.

\textit{Why are the microtubule clusters mostly central?}  An intriguing feature of microtubule-neurofilament segregation, which is consistent across all published reports, is that the microtubules generally cluster in the center of the axon, surrounded by a peripheral band of neurofilaments (see Introduction and Fig. \ref{fig_exp}).  It is interesting to note that this was usually the case in our simulations also.  According to our model, the segregation generated by microtubule clustering is caused by an exclusion of neurofilaments from the microtubule domain due to their failure to interact.  The central location of the microtubule bundle is essentially a boundary effect which arises because microtubules at the periphery of the axon can only be pulled towards microtubules that are located more centrally whereas microtubules in the center can be pulled towards microtubules on all sides.  The net result is that microtubule zippering by moving organelles tends to pull these polymers towards the axon center, displacing the neurofilaments to the periphery.  The organelles co-segregate with the microtubules because they must follow the available tracks.  

\textit{Segregation proceeds via the merging of small microtubule clusters.}  An interesting observation in our simulations is that microtubule-neurofilament segregation tends to proceed initially via the formation of small microtubule clusters that subsequently merge together. This was more apparent in simulations with smaller organelles, which are less efficient at zippering microtubules together (see discussion below).  Multiple small microtubule clusters have been reported in some studies on microtubule-neurofilament segregation induced by IDPN \cite {Papasozomenos1985, Zhu1998} (see Fig. \ref{fig_intermediate}D), but there is no published time course of segregation so it remains to be proven that these clusters are indeed intermediate states.  Interestingly, microtubule zippering in our simulations also gives rise to the formation of small microtubule clusters in healthy axons.  However, with ongoing neurofilament transport these clusters are transient and rarely merge to form larger ones.  This is consistent with reports that small clusters of microtubules, often adjacent to one or more membranous organelles, are commonly observed in electron micrographs of axons \cite {Price1991,Friede1970, Hirokawa1982, Wortman2014}.

\textit{Factors influencing rate of segregation.}  Our analysis gives us some insights into the factors that influence the rate of microtubule-neurofilament segregation.  First, given the same number density, larger organelles are more effective at causing segregation, because they can interact with more microtubules simultaneously and they can pull together microtubules that are farther apart.  Second, segregation occurs faster if the flux rate of the organelles is larger.  Third, segregation occurs faster if the degree of neurofilament transport impairment is larger.  These predictions are experimentally testable.  It is also clear that there must be some dependence on the density of motors on the organelle surface, as well as the neurofilament:microtubule ratio.  We are currently performing an extensive investigation on how the segregation phenomena depend on combinations of the model parameters using model simplification, nondimensionalization and mathematical analysis. These efforts will provide further insight of the biological problem and will be published elsewhere in the future.

\textit{The predicted rate of segregation is comparable to that in real axons.}  The best experimental data on the kinetics of microtubule-neurofilament segregation is for animals treated with the neurotoxin IDPN.  However, the rate of segregation in those animals depends on the mode of administration. When applied systemically to rats using a single intraperitoneal injection, segregation was first noted after 4 days, and after 4 such injections at 3 day intervals, the resulting segregation persisted for 6-16 weeks \cite {Papasozomenos1981}.  In contrast, when applied locally at high concentration by sub-perineurial injection into peripheral nerve, microtubule-neurofilament segregation was evident after 2 hours, with the microtubule clusters becoming increasingly compact over the next 4-10 hours \cite{Griffin1983,Griffin1983CDI}.  Nagele et al. \cite {Nagele1988a} analyzed the pairwise distance between microtubules (PDMT) and observed full compaction by 8 hours after injection. Sixteen hours later, segregation was no longer seen in most axons, indicating an almost complete reversal \cite{Griffin1983CDI}.  In our simulations, we observed segregation within 4-12 hours of a complete cessation of neurofilament transport, and remixing within 2-8 hours after a complete resumption of neurofilament transport.  This rate of segregation is comparable to the kinetics observed experimentally for injections of IDPN into nerves, and suggests that this delivery method results in a transient but acute inhibition of neurofilament transport.  We predict that the slower time course of segregation that is observed when IDPN is administered systemically is due to the lower effective dose experienced by the axons in those studies.  The rate of remixing was a bit shorter in our simulations than in the experimental reports, which may be because we assumed an instantaneous recovery of neurofilament transport rather than a gradual one, which is more likely.

\textit{What is the mechanism of neurofilament transport impairment?}   It is important to note that the impairment of neurofilament transport that leads to microtubule-neurofilament segregation in toxic neuropathies and neurodegenerative diseases also leads eventually to focal neurofilament accumulations and axonal swellings (see Introduction). Since microtubules are the tracks along which neurofilaments move, and since microtubule-neurofilament segregation appears early and precedes neurofilament accumulation  and axonal swelling, it has been hypothesized that the segregation reflects the uncoupling of neurofilaments from their transport machinery  \cite {Griffin1983,Llorens2013}.  Our modeling supports this hypothesis, but the molecular mechanism is unclear.  Many of the neurotoxic agents that cause microtubule-neurofilament segregation and impair neurofilament transport (e.g. hexanedione, IDPN, carbon disulfide) are reactive molecules that could, or are known to, modify neurofilaments chemically \cite {Llorens2013}.   It is thought that these compounds react with specific amino acid residues to form protein adducts which may then modify protein interactions, and that such chemical modifications target neurofilaments preferentially or that they somehow render these polymers more susceptible than other cargoes to transport impairments.  This selectivity could arise, for example, due to the unique structure or unusual amino acid composition of neurofilament proteins.  The mechanism of impairment could be by interfering with their interaction with molecular motors or with the interaction of these motors with the microtubule tracks. Future experimental studies will be required to resolve such questions.

\textit{How do neurofilament accumulations arise?}  The mechanism by which neurofilament accumulations arise is also of great interest given that this occurs in so many neurodegenerative diseases. Since local accumulations can only form if more neurofilaments move into a segment of axon than move out, the appearance of local swellings along axons implies some longitudinal instabilities in the transport of these cargoes.  Therefore we propose that neurofilament segregation is an early event in neurofilament transport impairment but that longitudinal instabilities or non-uniformities in the transport impairment must arise to give rise to local accumulations and axonal swellings.  We plan to address this in future studies.  Due the complex spatial and temporal nature of this problem, which entails the interactions of multiple dynamic components, we believe that a full understanding can only be achieved by a combination of experimental and modeling approaches.  Our present study is an important first step.
 
 \section*{Acknowledgments}
This project is supported by the NSF  DMS1312966 to CX. CX is also supported by the Mathematical Biosciences Institute (MBI) at the Ohio State University as a long-term visitor. BS was supported by MBI via a postdoc fellowship and in part by NSF DMS 1358932. AB is supported by NIH R01 NS038526. 
 
\section*{Supporting Information A: calculation methods for the RDF and OPD in Fig. 3}

To calculate the RDF, for each time frame we used the centers of the neurofilaments in the domain $ [-0.2  \mu \rm{m}, $ $ 0.2 \mu \rm{m}] \times [-0.2 \mu \rm{m}, 0.2 \mu \rm{m}]$ as reference points. We chose to use reference points in a smaller domain in order  to avoid  boundary effects. For each reference point we binned the center-to-center distances ($r$) between the reference neurofilament and all other neurofilaments with a bin size $\Delta r =  1 \rm{nm}$, and normalized by the factor $2\pi r \Delta r \rho^N$ where $\rho^N$ is the density of neurofilaments in the whole domain. We then took the average over all reference points for each time frame and then over all time frames to obtain $g(r)$.  We note that $g(r)$ is noisy  due to under sampling and the small bin size (1 nm). Averaging over more time frames or using a larger bin size gave smoother $g(r)$.  To compute the OPD $p_n$, we used circular windows with fixed radii 60 nm as in \cite{Kumar2002}. For each time frame, we sampled centers of $10 n^N $ circles according to the uniform distribution in the domain $[-0.3  \mu \rm{m}, 0.3 \mu \rm{m}] \times [-0.3 \mu \rm{m}, 0.3 \mu \rm{m}]$ and calculated the particle occupancy number for each circle. We then produced a histogram of occupancy numbers obtained for all circles and normalized it by the total number of circles to obtain $p_n$.
  
\section*{Supporting Information B: parameter estimation}
\label{sec_params}

The parameters of the model are summarized in Table 1.   In the following, we explain the methods used to estimate the parameters. 
 
\subsection*{Kinetic rates for neurofilament and organelle transports. \, }  
We assume that a neurofilament can only bind to a single microtubule at one time whereas an organelle, which is much larger, can interact simultaneously with one or more microtubules. The capturing radius $R_b$ for these interactions is taken to be 80 nm, which is the length of a kinesin motor when fully extended, as measured by electron microscopy \cite{Hirokawa1989}. 

The movement of neurofilaments along axons has been modeled as a bidirectional independent velocity jump process in 1D with two distinct pausing states in which a neurofilament either moves anterogradely or retrogradely, pauses for a short time, or pauses for a long time  \cite{Li2014,Monsma2014}.  Since that model is one dimensional, the kinetic rates extracted represent space averages over all neurofilaments in the axonal cross-section. For this reason, the kinetic rates for neurofilament transport in our model are related, but not identical, to the rates extracted in those other studies. We estimate $k_{on}^N$, the rate at which a neurofilament binds to a nearby microtubule within distance $R_b$ in our model,  to be five times as large as the transition rate $\gamma_{01}$,  the rate for a short-pause neurofilament to start moving,  defined in \cite{Monsma2014}. We estimate $k_{off}^N$,  the unbinding rate of neurofilaments in our model,  to be the same as the rate $\gamma_{10}$,  the rate for a moving neurofilament to stop and pause,  as defined in \cite{Monsma2014}.  Using $\gamma_{01} = 2.0\times 10^{-3} / \rms $ and $\gamma_{10} = 6.5 \times 10^{-2} / \rms$ for myelinated axons, we obtain $k_{on}^N = 1.0 \times 10^{-2} / \rms$ and $k_{off}^N = 6.5 \times 10^{-2} / \rms$. 

The rate that a neurofilament leaves the domain,  $k_{out}^N$,  is estimated in the following way. The average time for a neurofilament to move through $D$ is $L_N/s^N$, where $L_N$ is the average length of moving neurofilaments and $s^N$ is the  speed of the filament. Assuming that the neurofilament departure events are exponentially distributed then the rate $k_{out}^N$  is the reciprocal of the average time, i.e., $k_{out}^N = s^N/L_N$. Taking $L_N \approx 5 \mu\mathrm{m}$, which is the approximate average length of moving neurofilaments in cultured neurons, and  $s^N \approx 0.5 \mu \mathrm{m/s}$ \cite{Wang2000,Wang2001}, we obtain $k_{out}^N = 0.1 \, \rm{s}^{-1}$. 
 
The binding and unbinding rates of organelles to a nearby microtubule, $k_{on}^O$ and $k_{off}^O$, are assumed to be constant $2/\rms$ which are comparable to the rates used in previous mathematical models of vesicular transport along microtubules \cite{Kunwar2008,Erickson2011}. 

The passage rate for the organelles is calculated using their cross-sectional density. Assuming the cross-sectional organelle density is $\rho^O$, then the total number of organelles in $D$ is given by $\pi R_0^2\rho^O $. Denoting the speed of the organelle along microtubule to be $s^O$ and the length of the organelle to be $2a$, then the time that an organelle remains in $D$ is given by $2a/s^O$. Assuming that organelle arrival is a Poisson process, then the Poisson rate can be calculated as $ \pi R_0^2\rho^O$ divided by  $2a/s^O$, that is,  $$k_{in}^O =  \pi R_0^2 \rho^O s^O /(2a).$$ 

The sizes and densities of axonal organelles have been most carefully studied in \cite{Smith1980}. According to Table 1 in \cite{Smith1980}, there are two major sizes of organelles: large ones such as mitochondria have an average cross-sectional diameter of 280 nm, and small ones classified as tubular and vesicular profiles have an average cross-sectional diameter of 50 nm. The densities of these organelles were counted in longitudinal slices of axons, and need to be converted to cross-sectional densities for our model. We note that the cross-sectional density ($\rho^O$) and the longitudinal density (denote as $\rho^O_l$) are generally different. However, the area fraction occupied by the organelles averaged over cross sections and longitudinal sections are comparable, and this relation can be used to  convert the longitudinal density to cross-sectional density. Assuming that the objects are cylinders with radius $r$ and length $l$, the area fractions measured in cross and longitudinal sections are given by $\pi r^2 \rho^O$ and $2rl \rho^O_l$ respectively. Equating these two, we obtain $\rho^O = 2l \rho^O_l / (\pi r)$. For the large organelles, $2r = 280 \rm{nm}$, $l = 870 \rm{nm}$, $\rho^O_l = 0.023 / \micron^2$, and this leads to $\rho^O = 0.091  / \micron^2$. For the small organelles, $2r = 50 \rm{nm}$, $l = 180 \rm{nm}$, $\rho^O_l = 0.02 / \micron^2$, and this leads to $\rho^O = 0.0917 / \micron^2$.  

The density of organelles estimated above is comparable to the density data in published IDPN studies. For example, in Table 1 of \cite{Papasozomenos1981}, the mitochondrial density was measured to be $0.187-0.250 / \micron^2$. Taking $\rho^O = 0.187/\micron^2$, $a = 10b = 2.8 \micron $, $s^O = 1 \micron/\rm{s}$, and $R_0 = 1\micron$, we obtain $k_{in}^O =  \pi R_0^2 \rho^O s^O /(2a) = 0.105/\rms $.

\subsection*{Parameters in the pairwise repulsion and elastic spring forces. \,}   
We first estimate $L_r$ which is the maximum distance for pairwise repulsions of particles. The parameter $L_r$ for neurofilament-neurofilament repulsion is approximately $2 l^N$, where $l^N$ is the equilibrium brush thickness of the neurofilament sidearms. This was given by Eqn. (4) in   \cite{Kumar2002}, that is, 
\beq\tag{S1}
l^N = \left(\frac{12}{\pi^2}\right)^{1/3} N a \left( \frac{a}{s} \right)^{2/3}. 
\eeq
Here $N $ is the number of amino acids per sidearm, $a$ is the length of an amino acid and $s$ is the spacing between neurofilament sidearms. Taking $N = 679$ \cite{Kumar2002}, $a =3.5$ angstroms  \cite{Malacinski2005,Mingarro2000}, and $s = 3\, \rm{nm}$ \cite{Kumar2002,Geisler1981},  we obtain $l^N \approx 60.6 \, \rm{nm}$ and thus $L_r = 121.2 \, \rm{nm}$.  For simplicity, we use the same $L_r$ for pairwise interactions of all kinds of particles and their interactions with the domain boundary.  

We estimate the force prefactors $\eps^{kl}$ in the following way. We assume that the repulsion force between two neurofilaments is approximately 1 pN when their surface distance is $d = 40 \,\rm{nm}$. Under this assumption we have $\eps^{NN} = 1/(L_r/d-1)$. Taking $L_r = 121.2 \, \mathrm{nm}$, we obtain $\eps^{NN} \approx 0.5\,\rm{pN}$. We denote $\eps^{NN}$ by $\eps_r$ for simplicity of notation. We assume that the force prefactor for microtubule-microtubule, microtubule-neurofilament repulsions and repulsions of microtubules and neurofilaments with the boundary are the same as $\eps_r$. For repulsions that involve organelles, we use a prefactor that is five times as large, that is, $\eps^{kO} = \eps^{Ok} = 5 \eps_r$. 

Organelle movement can cause significant flow of the axoplasm near their surfaces and displace nearby microtubules and neurofilaments. As an organelle pushes into $D$, its radius increases and it pushes nearby fluid and particles away from itself;  as it moves away from $D$, instead of leaving void behind it, it creates negative pressure which draws the axoplasm to flow back and fill the space. In this model, we do not include the hydrodynamic interactions among these particles explicitly, but include this effect by adjusting the force factors $\eps$ associated with organelles. Specifically, when organelles push into the domain, we double $\eps^{kO}$ and $\eps^{Ok}$  to take into account the contribution of the fluid flow.   
 
The effective spring constants $\kappa^N$ and $\kappa^O$ are calculated in the following way.  The mean surface distance between a microtubule and a cargo engaged on it has been observed to be 17 nm (denote as $l_0$) \cite{Miller1985,Kerssemakers2006}.  We assume that the spring force and the repulsive force between a microtubule-neurofilament pair equilibrates at $l_0$, i.e.,  $\kappa^N l_0 = \eps_r (L_r / l_0 - 1) $. From this assumption we obtain 
\beq \tag{S2}
\kappa^N  = \eps_r  (L_r/l_0 - 1) /l_0. \label{def_kappaN}
\eeq
Plugging the values of $\eps_r$, $L_r$ and $l_0$ into this expression, we obtain $\kappa^N = 0.18 \, \rm{pN}/\rm{nm}$. 
Similarly, we assume that the spring force and the repulsive force of a microtubule-organelle pair equilibrates at $l_0$, and this leads to 
\beq \tag{S3}
\kappa^O  = 5 \eps_r (L_r/l_0 - 1) /l_0,  \label{def_kappaF}
\eeq
which is $\kappa^O = 0.9 \, \rm{pN}/\rm{nm}$. 
We note that the spring constant for a single kinesin motor is estimated to be $0.2-0.4 \, \mathrm{pN}/ \mathrm{nm}$ in \cite{Coppin1995}. The spring constants here are different from the spring constant of a single motor in two ways: first,  when a cargo moves along a microtubule there could be multiple motors being active, and thus the spring constants used here represent the sum of the spring constants over all active motors; second,  the spring constants used here only take into account projections of the elasticity of individual motors in the plane orthogonal to the microtubule, whereas molecular motors are most likely slanted when dragging a cargo along a microtubule.

\subsection*{Drag and diffusion coefficients}
  
We treat neurofilaments as slender cylinders, and estimate the drag coefficient per unit length ($\mu/L$) by the formula given in \cite{Cox1970},  $\mu^N/L = 4\pi\eta /(\ln(L_N/b) + \ln 2 -0.5)$, where $\eta$ is the viscosity of the axoplasm, $L_N$ is the characteristic length of the cylinder, and $b$ is the characteristic  radius of the cross-section.  The viscosity of the cell cytoplasm is estimated to be $\eta \sim  3-5\, \pN \cdot \rm{s} /\micron^{2}$ for mammalian cells \cite{Parhad1982,Swaminathan1997} \void{and $11\, \pN \cdot \rm{s} /\micron^{2}$ for bacteria \cite{Elowitz1999}}.  The persistence length of neurofilaments is 200-450 nm \cite{Beck2010,Wagner2007}. The characteristic radius of a neurofilament with its sidearms is about 20 nm \cite{Hsieh1994}. Taking $L_N =  500 \mathrm{nm}$ , $b = 20 \mathrm{nm}$,  we obtain $\mu^N/L =  4\pi\eta/(\ln(L_N/b) + \ln 2 -0.5) \approx 14.7 \,\pN \cdot \rm{s} /\micron^{2}$. Taking the length of the neurofilament to be $L = 5 \micron$, we obtain $\mu^N \approx 73.5 \, \pN \cdot \rm{s} / \micron$. 

We estimate the drag coefficient for microtubule in a similar way. The persistence length of microtubules has been estimated to be $80 \pm 20 \mu \mathrm{m}$ \cite{VandenHeuvel2008}. We estimate the characteristic radius of the microtubule with its associated proteins in axons to be 37.5 nm. Taking $\eta =  4\, \pN \cdot \rm{s} /\micron^{2}$, $L_M = 80\, \micron$ and $b = 37.5 \, \mathrm{nm}$, we obtain $\mu^M / L= 6.4 \, \pN \cdot \rm{s} /\micron^{2}$. We note that this estimate is close to the estimate obtained by treating microtubules as infinitely long cylinders, for which one can use the Oseen drag formula $\mu^M  / L = 4 \pi \eta / \log(4/R_e - \gamma + 0.5)$ where $\gamma \approx 0.5772$ is the Euler's constant  and $R_e$ is the Reynolds number  \cite{Oseen1910,Chwang1976}.  The characteristic flow velocity in axonal cross-section is approximately $U = 0.2 \mu \mathrm{m} / \mathrm{s}$ given by the pushing of the organelles. Assuming that the density of the cytoplasm $\rho$ to be the same as water, we find that $R_e = 2 U b \rho / \eta \approx 4.3\times 10^{-3}$. Plugging $\eta$ and $R_e$ into the Oseen's formula, we obtain the drag per unit length to be  $6.3\,\pN \cdot \rm{s} /\micron^{2}$.  Taking the sectional length of microtubule to be $L  = 80 \micron$, we obtain $\mu^M  \approx 6.4\times 80 = 512 \, \pN \cdot \rm{s}/\micron$. 

We estimate the drag coefficient for organelles using the formulas for a prolate ellipsoid given in \cite{Chwang1976}.  Assume that the major axis of the ellipsoid is $2a$ and the minor axis is $2b$, then the drag coefficient per unit length is given by 
\beqn
\mu^O/L =  \frac{16\pi\eta e^3}{2e + (3e^2 -1) \ln[(1+e)/(1-e)]} \ffor 0<e<1,  
\eeqn 
where $e = \sqrt{(1-(b/a)^2)}$ is the eccentricity.  For  $b/a = 0.1, 0.2 \mbox{ and } 0.5$,  we have $\mu^O/L \approx 14.4$, $17.9$ and 26.0 $ \pN \cdot \rm{s} /\micron^{2}$.  For organelles with $b = 140 \rm{nm}$ and $b/a = 0.1$, we have $\mu^O \approx 40.3 \rm{pN}\cdot \rm{sec}/\micron$.  

We calculate the diffusion coefficients $D_k$ with $k = M, N,$ and $O$ using the Einstein relation 
\beq\tag{S4}
D_k = k_B T/\mu_k, \label{Einstein}
\eeq
where $k_B$ is the Boltzmann's constant and $T$ is the absolute temperature. At room temperature ($25 \,^{\circ}{\rm C}$ or 298 K), one has $k_B T = 4.11 \rm{pN}\cdot \rm{nm}$. Using $\mu_N = 73.5  \, \mathrm{pN} \, \mathrm{s} \, / \mu \mathrm{m} $, we obtain $D_N \approx 5.6\times 10^{-5} \, \micron^2/\rm{s}\,$. We calculated $D_M$ and $D_O$ in a similar way. Finally $\sigma_k$   is given by  the relation  $\sigma_k = \sqrt{2D_k}$.

\section*{Supporting Information C: simulation algorithm}
\label{sec_algorithm}

To solve the model numerically, we chose a time step $h$ much smaller than all the time scales involved in Mechanisms 1-3, treated
the binding and unbinding, arrival and departure of cargoes explicitly at discrete time steps, and integrated the model system (Eqn. 8 in main text) using the explicit Euler's method. Because $\sigma_k$, $k = M, N, C$ are constant in time, the numerical integrator has  strong order 1.0  \cite{KloedenPlatenBook}.  For the segregation simulations, we used $h = 1/50 \sec$ if there was no organelle in $D$; otherwise we used $h = 1/1600 \sec$ in order to deal with the  stiffness of the equations introduced by the pushing of organelles when they move into $D$.  The algorithm for a typical time step is summarized below.
 
 \bigskip
{\noindent \bf Algorithm for the model}
\begin{enumerate}

\item Stochastic removal of moving neurofilaments. For each moving neurofilament, generate a random number $r$ uniformly distributed in [0,  1]. If $r <  1 - e^{-k_{out}^N h }$ then remove it from $D$. 

\item Update $z_i^O$  and $r_i^O$ for each organelle in $D$ according to Eqn. 1 and Eqn. 4 in the main text.  If $z_i^O$ becomes bigger than or equal to $a$ then remove the {\em i}-th organelle and release all microtubules from it.  

\item  Stochastic unbinding of cargoes, i.e., neurofilaments and organelles, from their engaged microtubules. If a microtubule and an organelle are engaged, then generate a random number $r$ uniformly distributed in [0 1],  and let them unbind if $r< 1 - e^{-k_{off}^O h }$. They also unbind if their surface distance is bigger than $R_{b}$. The same method is used to update microtubule-neurofilament binding. 

\item Stochastic binding of cargoes to microtubules. 
\begin{enumerate}
\item If the surface distance of a cargo and a microtubule is smaller than $R_{b}$, then generate a random number and determine if they intend to bind to each other. Loop through all microtubule-neurofilament  and microtubule-organelle pairs, and find all potential binding events. 
\item Accept or reject the potential binding events according to the availability of the associated microtubules in a random order. We assume that one microtubule has 5 tracks, each neurofilament occupies one track, an organelle with maximum radius $140 \rm{nm}$ occupies 2 tracks, and an organelle with maximum radius $70 \rm{nm}$ occupies 1.5 tracks. 
\end{enumerate}

\item  Addition of new neurofilaments and organelles to $D$. The number of new neurofilaments equals the number that has been removed in step 1. The number of new organelles is determined using the rate $k_{in}^O$. Since the time step $h$ is much smaller than the mean arrival time of organelles $k_{in}^O$, we introduce a new organelle in each time step with probability $(1-e^{-k_{in}^O h})$. We then add these cargoes at $l_0=17 \, \rm{nm}$ away from a randomly chosen microtubule at a random angle. If the random location overlaps with an existing particle or the associated microtubule is too crowded, then a different microtubule and an angle is generated randomly. Once a new cargo is added, it is bound to the selected microtubule.   

\item Update the positions of the microtubules, organelles  and neurofilaments in $D$ by integrating the model system (Eqn. 8 in the main text) using the explicit Euler's method, i.e., 
\beq\tag{S5}
\begin{aligned}
& \bx^k_i (t + h)  = \bx^k_i (t ) +  \bF_i^k(t)  h  /\mu^k + \sigma_k \bold{r}^k_i \sqrt{h} , \\ 
& \qquad 1 \leq i  \leq  n^k,  \quad k = M, N, F.
\end{aligned}
\eeq
where $\bold{r}^k_i$ are pairs of random numbers generated from the standard 
normal distribution using the Box-Muller transform  \cite{Box1958}.  To avoid large values of $\bold{r}^k_i$, they are regenerated if the absolute value of any component is greater than 5.  

\item Go to Step 1 for the next time step.

\end{enumerate}


%
%
%


\end{document}